\documentclass[pdflatex,sn-mathphys,iicol]{sn-jnl}
\jyear{2023}

\usepackage{graphicx}
\usepackage{subfig}
\usepackage{enumerate}
\usepackage{amsmath,mathtools}
\usepackage{lineno}
\usepackage[numbers]{natbib}
\usepackage{amsfonts,bm}
\usepackage[commentmarkup=uwave]{changes}

\definechangesauthor[name={Reviewer 1}, color = red]{R1}
\definechangesauthor[name={Reviewer 2}, color = blue]{R2}
\definechangesauthor[name={Author}, color = orange]{Author}

\begin{document}

\title[Article Title]{Inference of relative permeability curves in reservoir rocks with ensemble Kalman method}

\author[1]{\fnm{Xu-Hui} \sur{Zhou}}\email{xuhuizhou@vt.edu}

\author[2]{\fnm{Haochen} \sur{Wang}}

\author*[3]{\fnm{James} \sur{McClure}}\email{mcclurej@vt.edu}

\author*[4]{\fnm{Cheng} \sur{Chen}}\email{cchen6@stevens.edu}

\author*[5]{\fnm{Heng} \sur{Xiao}}\email{heng.xiao@simtech.uni-stuttgart.de}

\affil[1]{\orgdiv{Kevin T. Crofton Department of Aerospace and Ocean Engineering}, \orgname{Virginia Tech}, \orgaddress{\city{Blacksburg}, \postcode{24060}, \state{Virginia}, \country{USA}}}

\affil[2]{\orgdiv{Thomas Lord Department of Mechanical Engineering and Materials Science}, \orgname{Duke University}, \orgaddress{\city{Durham}, \postcode{27708}, \state{North Carolina}, \country{USA}}}

\affil*[3]{\orgdiv{National Security Institute}, \orgname{Virginia Tech}, \orgaddress{\city{Blacksburg}, \postcode{24060}, \state{Virginia}, \country{USA}}}

\affil*[4]{\orgdiv{Department of Civil, Environmental and Ocean Engineering}, \orgname{Stevens Institute of Technology}, \orgaddress{\city{Hoboken}, \postcode{07030}, \state{New Jersey}, \country{USA}}}

\affil*[5]{\orgdiv{Stuttgart Center for Simulation Science}, \orgname{University of Stuttgart}, \orgaddress{\city{Stuttgart}, \postcode{70569}, \state{Baden-Württemberg}, \country{Germany}}}

\abstract{
Multiphase flows through reservoir rocks are a universal and complex phenomenon. 
Relative permeability is one of the primary determinants in reservoir performance calculations. Accurate estimation of the relative permeability is crucial for reservoir management and future production.
In this paper, we propose inferring relative permeability curves from sparse saturation data with an ensemble Kalman method. We represent these curves through a series of positive increments of relative permeability at specified saturation values, which guarantees monotonicity within, and boundedness between, 0 and 1. The proposed method is validated by the inference performances in two synthetic benchmarks designed by SPE and a field-scale model developed by Equinor that includes certain real-field features. The results indicate that the relative permeability curves can be accurately estimated within the saturation intervals having available observations and appropriately extrapolated to the remaining saturations by virtue of the embedded constraints. The predicted well responses are comparable to the ground truths, even though they are not included as the observation. The study demonstrates the feasibility of using ensemble Kalman method to infer relative permeability curves from saturation data, which can aid in the predictions of multiphase flow and reservoir production.}


\maketitle

\section{Introduction}\label{sec1}
Multiphase flows in porous media occur in a wide range of engineering applications. In water treatment, porous materials are used to remove bacteria and harmful substances from the water supply~\cite{zhang2017superwetting,cheng2020multifaceted}; in chemical engineering, packed bed reactors are employed to facilitate the heterogeneous reactions~\cite{parker1989multiphase,noorman2007packed}; and in the mitigation of global warming, supercritical carbon dioxide is injected into porous rock formations for geologic carbon storage~\cite{moodie2021relative,bickle2009geological}. These examples and many other applications demonstrate the significance of understanding and modeling the multiphase flows in porous media for accurate prediction of the system performance and efficient operation. 

Historically, the development of models for multiphase flow has largely been driven by the petroleum engineering industry, with the goal of achieving more efficient oil and gas recovery from hydrocarbon reservoirs~\cite{parker1989multiphase}. In petroleum industry, oil is displaced and driven to the production wellbore by injecting water or gas in the secondary EOR (Enhanced Oil Recovery)~\cite{adibifard2018novel}. In this process, the relative permeability plays a crucial role in determining the motion of multiphase flow, as it reflects the ability to transmit a particular fluid in the presence of other immiscible fluids. Inaccurate estimation of the relative permeability may result in reduced oil production and other problems~\cite{honarpour2018relative}. 
Therefore, it is essential to accurately determine the relative permeability in order to minimize uncertainties in reservoir management and mitigate the impact of inaccurate estimations on oil production.

Relative permeability curves are usually obtained through steady- or unsteady-state core flooding experiments~\cite{henderson1998measurement,pini2013simultaneous,fan2019comprehensive,fan2020influence}. The steady-state methods have the highest accuracy since the capillary equilibrium is achieved in the measurements, enabling direct calculation of the effective permeability for each phase at a given saturation using Darcy's law. However, measurement for each saturation takes hours or days, making the steady-state methods inherently time-consuming and expensive. The unsteady-state measurements are more prevalent because they do not require the equilibrium that a set of relative permeability curves can be obtained within only a few hours. But they are less reliable and considered only as qualitative substitutes of the steady-state measurements. This is because the relative permeability data is obtained at an unsteady state with changing properties at the estimated saturation, which may lead to very different values from the tests performed earlier. Due to the complexity and costs in laboratory measurement, a number of empirical models have been developed to estimate the relative permeability values~\cite{corey1954interrelation,brooks1966properties,honarpour2018relative,chierici1984novel,alpak1999validation,stone1970probability,stone1973estimation}. Among these, the most well-known is the Corey's model in power law relations~\cite{corey1954interrelation,brooks1966properties}, which is a theoretical approach developed from the Burdine equations~\cite{burdine1953relative}. Other models, such as the Hornarpour model~\cite{honarpour2018relative}, are more based on real measurements and developed for different reservoir conditions. 

Laboratory measurements of relative permeability curves have a major shortcoming in that they may not be able to accurately describe the multiphase flows in field-scale reservoirs~\cite{matthews2008using}.
This is due to the vast disparity in spatial scales between the core samples in laboratory tests and industrial reservoirs, as well as their different reservoir conditions~\cite{honarpour2018relative,fanchi2005principles}. For a field-scale reservoir, history matching the production data is a typical way for estimating the relative permeability curves. Mathematically, calculating parameters (relative permeability) from observation data (production data) is an inverse problem, which can be solved with optimization methods. For example, 
Reynolds et al.~\cite{li2001simultaneous} used the adjoint method to estimate the three-phase oil relative permeability curve by analyzing the sensitivity of production data to parameters defining the relative permeability functions. To derive the curve, they relied on two sets of two-phase relative permeability curves (oil-gas and oil-water) based on Stone's Model II~\cite{stone1973estimation} and represented by power law models. Similarly, Eydinov et al.~\cite{eydinov2009simultaneous} estimated the relative permeability curve with the same method by history matching the three-phase flow production data. However, they represented the relative permeability curve with the B-splines, which was much more flexible than the power law models but required additional constraints for the coefficients to ensure the monotonicity. They also guaranteed the possible convex property of the curve by constraining its derivatives. 

Another approach is the ensemble-based method, which has shown promising performances in recent decades with applications to geoscience~\cite{ahmed2023frequency,wang2022benefit,fossum2022verification,cruz2022joint}, turbulent flow modeling~\cite{strofer2021dafi,zhang2022ensemble}, and  medical physics~\cite{rortveit2021reducing,naevdal2022fluid}, among others. It differs from the adjoint-based method in that the derivatives are not required but replaced by covariance matrices computed from the ensemble~\cite{kovachki2019ensemble}. This makes it more straightforward and less labor-intensive to implement, especially for complex systems such as the reservoirs. Li et al.~\cite{li2010ensemble} inferred the oil-water permeability curves by history matching the production data with the ensemble Kalman filter (EnKF). The curves were represented with the flexible B-spline model while the monotonicity was guaranteed by solving a system of linear equations for the coefficients. Despite the success in the above works, they were only tested on simple synthetic examples and not on realistic reservoirs. Other studies have demonstrated the applicability of EnKF to inferring relative permeability curves in real-world reservoirs~\cite{skjervheim2007incorporating,haugen2008history,bianco2007history,evensen2007using,seiler2009advanced,chen2010ensemble}. Seiler et al.~\cite{seiler2009advanced} showed that EnKF led to improved predictions and fast updating of the relative permeability curves on a complex North Sea oil field with the most recent production data. The Corey's model was employed to represent the relative permeability curves and the endpoint saturations were also inferred. Chen et al.~\cite{chen2010ensemble} applied the similar representation and optimization method to the Brugge field, but they included additional permeability vectors in the history matching process to assist the inference.

Most of the studies in this line seek to model relative permeability curves with a suitable representation and infer the curves from field data using adjoint- or ensemble-based methods. Power law models contain intrinsic monotonicity but may lack the complexity required for real-world fields, while B-spline models are more flexible but require additional constraints.
In this paper, we introduce a simple and novel representation for the relative permeability curve and use an ensemble Kalman method to infer the parameters therein based on sparse saturation measurements. The representation has the merits of embedded monotonicity and boundedness that are physically required for reservoir simulations. The proposed method is evaluated in a series of test cases consisting of two synthetic benchmarks and a more realistic project.
Most of the works on inferring relative permeability curves by history matching are performed using reservoir production data. In recent years, the increasing use of permanent sensors and advancements in 4D seismic monitoring have made it possible to collect a broader range of data from reservoir fields~\cite{maleki2022machine,sengupta2022cpet,corte2023bayesian,cruz2022joint}, which may assist in the estimation of relative permeability curves. In this study, we also investigate the inference capability by using production data only, and compare it to that based on sparse satuation data.

The rest of the paper is structured as follows. The challenges associated with modeling fluid flows in reservoir rocks and the commonly used black-oil model are described in Section~\ref{sec2}.
The representation of relative permeability curves and the ensemble-based inference are presented in Section~\ref{sec3}. The case-setup and inferring performance of the proposed method for three test cases are detailed in Section~\ref{sec4}. The paper is concluded in Section~\ref{sec5}.

\section{Flow in reservoir rocks}\label{sec2}
The flow of fluids in reservoir rocks is an intricately complex phenomenon that poses significant challenges to the oil and gas industry. The multiphase nature of these flows, combined with the presence of porous rock formations, dynamic pressure and temperature gradients, creates a dynamic and constantly evolving system that can be difficult to model and predict.

At the heart of this complexity is the mixture of oil, gas, and water that comprises the fluid phase within the reservoir. Each of these fluids has different physical properties and flow behavior. Specifically, the oil can exist in a variety of forms, ranging from liquid to gas, with the gas phase either free or dissolved within the oil phase. This can lead to a wide range of behaviors, including phase segregation, trapping, and mobilization, which must be accurately modeled in order to predict reservoir behavior. Additionally, the water phase, which can exist as either brine or freshwater, is also present in varying quantities and can further complicate the flow behavior of the fluid mixture. These factors, combined with the complex geometry and heterogeneity of reservoir rocks, make accurate modeling of fluid flows in these environments a significant challenge for the petroleum engineering community.

In addition to these inherent complexities, the oil and gas industry must also contend with the challenges of extracting hydrocarbons from these complex multiphase flow systems. This requires sophisticated modeling and simulation techniques, as well as a deep understanding of the physical properties of the fluids and the rock formations within the reservoir. These efforts are further complicated by the need to balance production rates against reservoir depletion, with production strategies often requiring constant adaptation in response to changing reservoir conditions.

The most widely used fluid model for reservoir simulations is the black-oil model, which represents the behavior of oil, gas, and water in the reservoir by a set of equations. The black-oil equations are deduced based on mass conservation for each component, along with Darcy's law and initial and boundary conditions for closure. Here we briefly recall the model equations for mass and momentum conservation,
\begin{equation*}
\begin{gathered}
\frac{\partial}{\partial t}\left(\phi_{\mathrm{ref}} A_\alpha\right)+\nabla \cdot \mathbf{u}_\alpha+q_\alpha=0, \\
\mathbf{v}_\alpha=-\frac{K_{r,\,\alpha}}{\mu_\alpha} \mathbf{K}\left(\nabla p_\alpha-\rho_\alpha \mathbf{g}\right),
\end{gathered}
\end{equation*}
where $\phi_{\mathrm{ref}}$ denotes reference porosity, $\mathbf{K}$ is permeability of the porous medium, and $\mathbf{g}$ is gravitational acceleration. For component $\alpha$, $A_\alpha$ and $q_\alpha$ denote its accumulation and well outflux density; $K_{r,\,\alpha}$, $\mu_\alpha$, and $p_\alpha$ denote its relative permeability, viscosity, and phase pressure; $\mathbf{u}_\alpha$ and $\mathbf{v}_\alpha$ are its component velocity and phase velocity, which are related by a shrinkage/expansion factor $b_\alpha$. Specifically, the accumulation terms and fluxes are calculated as
\begin{equation*}
\begin{array}{ll}
A_\textrm{w}=m_\phi b_\textrm{w} S_\textrm{w}, & \mathbf{u}_\textrm{w}=b_\textrm{w} \mathbf{v}_\textrm{w}, \\
A_\textrm{o}=m_\phi\left(b_\textrm{o} S_\textrm{o}+r_\textrm{og} b_\textrm{g} S_\textrm{g}\right), & \mathbf{u}_\textrm{o}=b_\textrm{o} \mathbf{v}_\textrm{o}+r_\textrm{og} b_\textrm{g} \mathbf{v}_\textrm{g}, \\
A_\textrm{g}=m_\phi\left(b_\textrm{g} S_\textrm{g}+r_\textrm{go} b_\textrm{o} S_\textrm{o}\right), & \mathbf{u}_\textrm{g}=b_\textrm{g} \mathbf{v}_\textrm{g}+r_\textrm{go} b_\textrm{o} \mathbf{v}_\textrm{o},
\end{array}
\end{equation*}
where $m_\phi$ is pore volume multiplier determined by pressure; $r_\textrm{og}$ and $r_\textrm{go}$ denote the ratio of vaporized oil to gas in gaseous phase and ratio of dissolved gas to oil in oleic phase, respectively; $S_\textrm{w}$, $S_\textrm{o}$, and $S_\textrm{g}$ denote the saturation of water, oil, and gas, respectively, with the relation of $S_\textrm{w} + S_\textrm{o} + S_\textrm{g} = 1$.
A complete description of the black-oil model equations can be found in the literature~\cite{rasmussen2021open}.

As is shown above, relative permeability $K_{r,\,\alpha}$ plays a critical role in determining the phase flux by Darcy's law since it describes the fractional flow of each fluid phase through the porous medium. Without reliable data on relative permeability, predictions of fluid behavior can be highly inaccurate. To address this problem, it is essential to obtain accurate relative permeability curves, which serve as a basis for developing reliable models of petroleum reservoirs and optimizing production and recovery strategies.

\section{Methodology}\label{sec3}
This work aims to demonstrate a data-driven framework for inferring relative permeability curves in reservoir rocks.
The inference needs to represent the relative permeability curves appropriately, and incorporate sparse and potentially noisy measurements in the reservoir.

\begin{figure*}[!htb]
\centering
\includegraphics[width=0.96\textwidth]{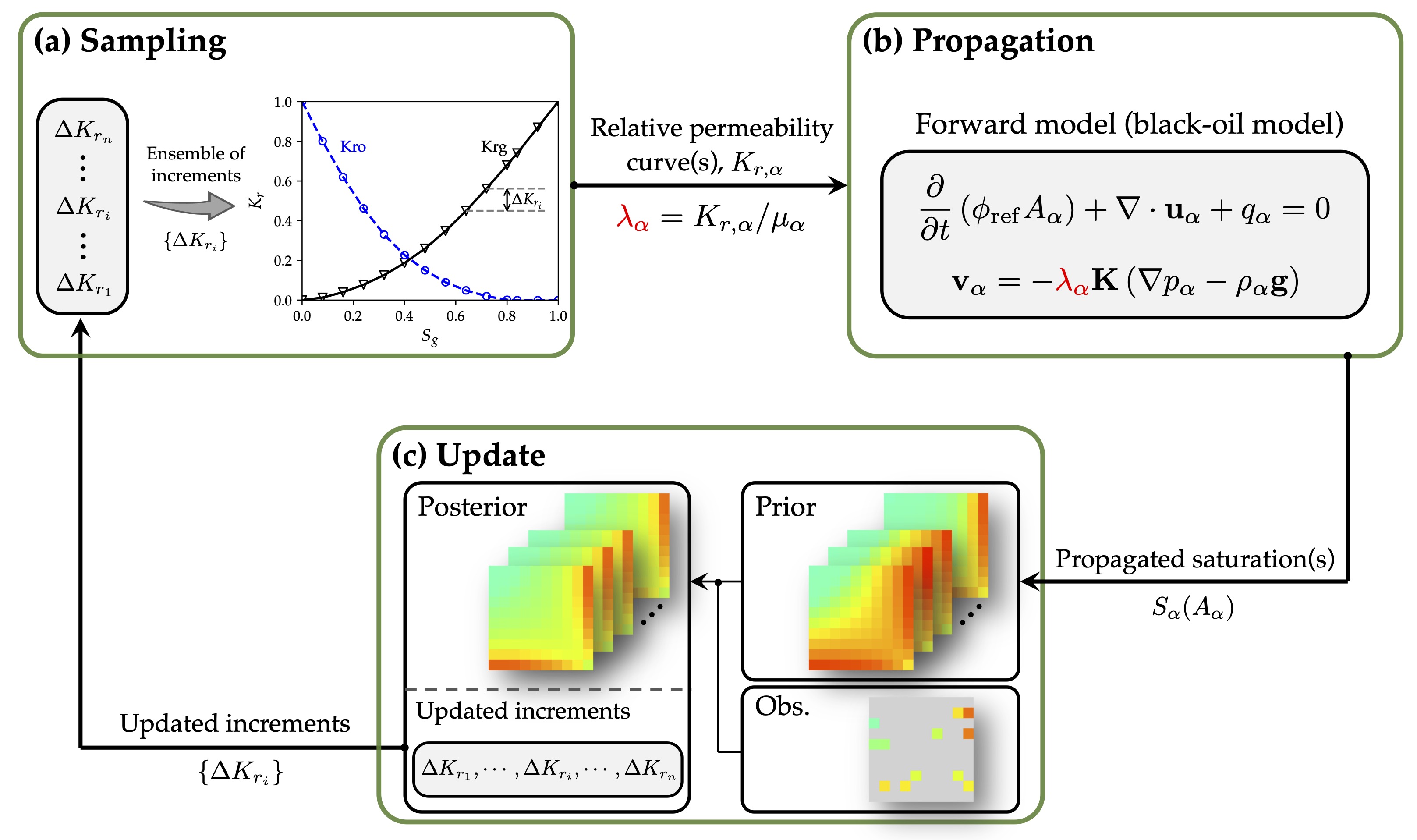}
  \caption{
  Schematic of the ensemble-based inference of relative permeability curve with sparse saturation data, consisting of three main steps: (a) sample the parameters to determine the increments $\left\{\Delta K_{r_i}\right\}$ for representing the curve, (b) propagate the represented curves to saturation fields by solving the black-oil equations, and (c) update the parameters by incorporating the sparse observation data. The observation data in this study is taken at a specific moment in time, which differs from many similar works that use time series data.}
  \label{fig:schematic}
\end{figure*}

We propose representing the curve by a collection of control points with increasing relative permeabilities, and using the ensemble Kalman method to infer the parameters (denoted as $\bm{\omega}$) therein based on the sparse observation data. The framework for inferring the relative permeability curves is presented in Fig.~\ref{fig:schematic}. It consists of three steps: (1) parameter sampling, (2) forward model propagation, and (3) parameter update with measurements, which are shown in Figs.~\ref{fig:schematic}a--c, respectively, and will be described in detail below.

\subsection{Relative permeability curve representation with hard constraints}
Parameterization of relative permeability curves is a crucial component of the inverse problem. The curve is in a rather simple form with only one dimension of saturation. However, it must satisfy the constraints as follows.
~\\
\begin{enumerate}[(i)]
  \item The curve is an increasing function of the corresponding saturation.
  \item The curve is bounded, with the lower bound of relative permeability being zero and the upper bound being one.
\end{enumerate}
~\\
A common method to represent such a simple function is using a truncated Chebyshev polynomials series~\cite{zhang2022assessment}. The coefficients of the polynomials are optimized to fit the available data. However, this representation cannot strictly guarantee the constraints throughout the optimization process. Modifying the coefficients alters the curve form globally and may induce non-monotonicity at some intervals, even the monotonicity constraint is softly imposed through a regularization term in the cost function (in Fig.~\ref{fig:schematic}c). The non-monotonic curve will cause a quick termination of the reservoir simulation and the parameter inference. A more generalized, neural-network-based representation may have a similar difficulty of not being able to ensure the monotonicity always.

\begin{figure}[!htb]
\centering
\includegraphics[height=0.26\textwidth]{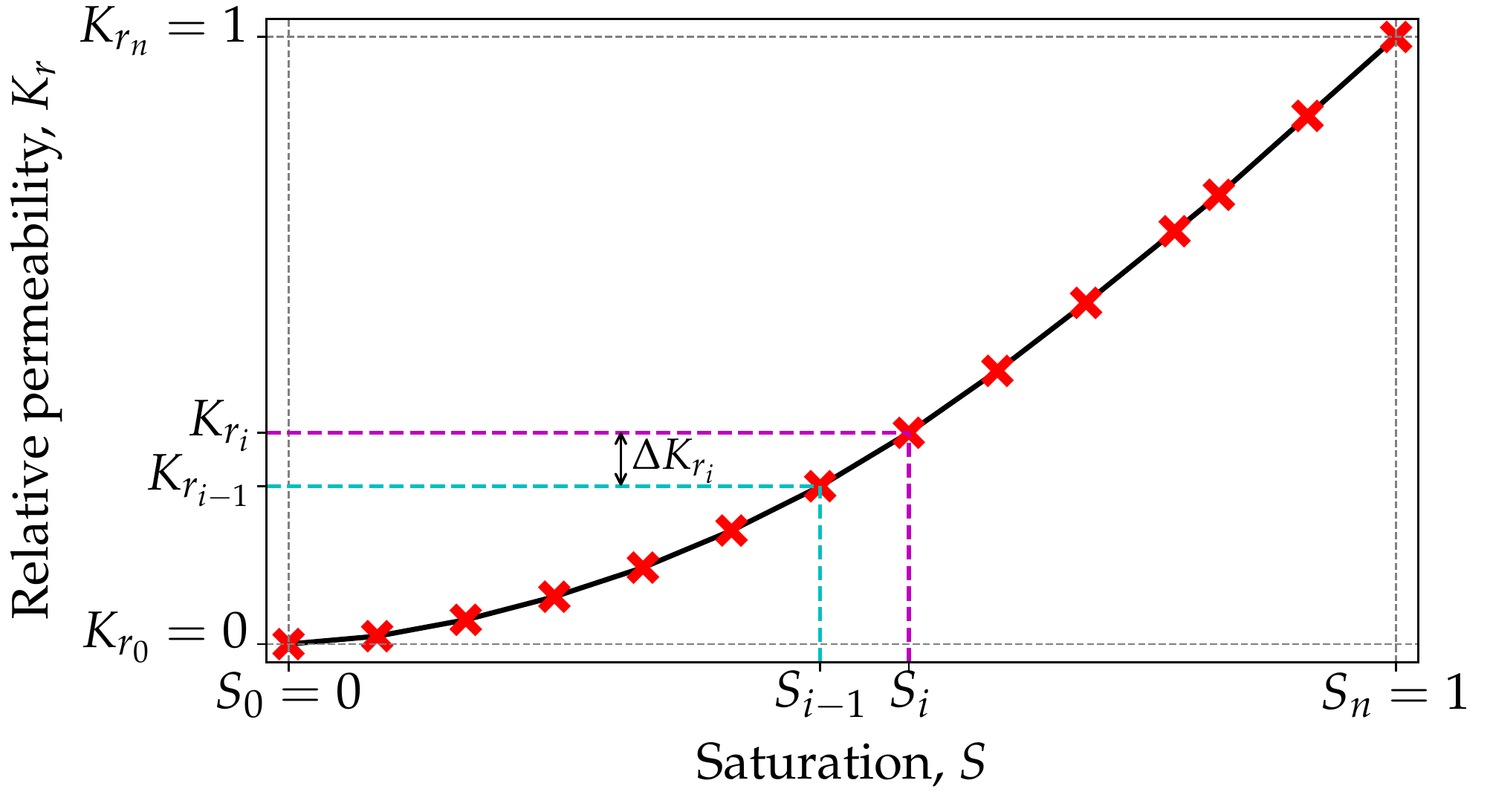}
  \caption{
  Representation of relative permeability curves using a set of control points with increasing relative permeabilities. The monotonicity is guaranteed by the positive increments in relative permeability ($\Delta K_r$) between any two contiguous points, while the boundedness is ensured by the starting and ending points.
  }
  \label{fig:representation}
\end{figure}

We propose representing relative permeability curves by a set of control points with increasing relative permeabilities, which is shown in Fig.~\ref{fig:representation}. Specifically, the curve is regulated by $n+1$ control points with the saturations $S_0,S_1, \ldots, S_n$ and corresponding relative permeabilities $K_{r_0},K_{r_1}, \ldots, K_{r_n}$. To satisfy the boundedness constraint, we assign the lower bound $K_{r_0} = 0$ and the upper bound $K_{r_n} = 1$ for $S_0 = 0$ and $S_n = 1$, respectively. Monotonic behavoir can be obtained when the increments in relative permeability between any two adjacent points are always positive. To this end, we represent the increments using the natural exponential functions. For example, for the neighboring saturations $S_{i-1}$ and $S_i$, we set the increment in relative permeability $\Delta K_{r_i} = \exp (\omega_i)$, where $\omega_i$ is a parameter to be inferred. However, such a determined upper bound is essentially the sum of all increments, i.e., $K_{r_n} = \sum_{i=1}^n \exp(\omega_i)$, which may surpass the upper bound of one defined above for the boundedness constraint. Here we are inspired by the normalization for feature scaling in machine learning, and normalize the increments by their sum, capping the maximum relative permeability ($K_{r_n}$) at one. As such, the representation embeds both monotonicity and boundedness to model the relative permeability curve and significantly reduces the risk of a black-oil simulation crash. The control points are determined as
~\\
\begin{small}
\begin{equation*}
K_{r} (S_i)= \begin{cases}0, & \text{if } i=0, \\ \sum_{m=1}^i \frac{\exp (\omega_m)}{\sum_{j=1}^n \exp (\omega_j)}, & \text{if } i=1, \ldots, n-1, \\ 1, & \text{if } i=n,\end{cases}
\end{equation*}
\end{small}
~\\
where $i$, $j$, $m$ denote the indexes of the parameter or control point.
The curve is then characterized by these control points and the line segments connecting adjacent points, and the parameters $\bm{\omega} = \left[\omega_1, \ldots, \omega_n\right]$ are optimized to accommodate the observation data.

Note that the number of control points and their saturation distribution can be chosen with flexibility. In this work, we select $n \leq 10$ and distribute the saturations uniformly (i.e., $S_i = i/n$).

\subsection{Ensemble-based inference for relative permeability curve}

With the appropriate representation for relative permeability curves, we employ the ensemble Kalman method to infer the parameters $\bm{\omega}$ based on the saturation data. The ensemble Kalman method is traditionally known as a data assimilation approach to estimate possible states of a system (e.g., saturation, pressure, and production in a reservoir) as it evolves in time. But here we infer the parameters that define relative permeability curves, which is performed uing the iterative ensemble Kalman method. The procedure is detailed as follows.
~\\
\begin{enumerate}[(i)]
  \item Sample the parameters $\bm{\omega}$ based on the initial prior distributions and represent the curve with the increments $\left\{\Delta K_{r_i}\right\}$ (Fig.~\ref{fig:schematic}a). Each increment is a function of all the parameters. The initial parameters, $\bm{\omega}^0$, are determined as $\omega_i^0 = 0.1$ for $i = 1, \ldots, n$ such that the curve is a straight line connecting $(0, 0)$ and $(1, 1)$. The initial ensemble is obtained by drawing random samples of parameters via the formula $\bm{\omega}_j = \bm{\omega}^0 + \bm{\epsilon}_j$, where $\bm{\epsilon} \sim \mathcal{N}(0, 0.5^2)$ and j denotes the index of the sample. Each of the sample $\bm{\omega}_j$ corresponds to a distinct relative permeability curve, generating an ensemble of curves for the reservoir simulation.
  \item Propagate each relative permeability curve in the ensemble to saturation fields by solving the black-oil equations (Fig.~\ref{fig:schematic}b).  The predicted observable quantities are obtained via post-processing from the simulated saturation fields, e.g., extracting oil/gas saturations at some locations in the reservoir.
  \item Update the parameters $\bm{\omega}$ based on the statistical analysis of the predicted observable quantities obtained in step (ii) and the comparison with the sparse observation data (Fig.~\ref{fig:schematic}c).
\end{enumerate}
~\\
Steps (ii) and (iii) are repeatedly executed until the convergence criterion is reached.

The cost function used in step (iii) is written as
\begin{equation}
J = \left\|\bm{\omega}_j^{l+1}-\bm{\omega}_j^l\right\|_\mathsf{P}^2 + \left\|\mathsf{y}_j-\mathcal{H}[\bm{\omega}_j^{l+1}]\right\|_{\mathsf{R}}^2,
\label{eq:cost-function}
\end{equation}
where $\|\cdot\|_\mathsf{A}$ denotes the norm weighted by the covariance matrix $\mathsf{A}$, $l$ is the iteration index, $\mathsf{y}$ is the observation, $\mathsf{P}$ is the model error covariance matrix indicating the parameter uncertainties, $\mathsf{R}$ is the observation error covariance matrix, and $\mathcal{H}$ is the observation operator. The first term in Eq.~(\ref{eq:cost-function}) is used to regularize the updated parameters by penalizing deviations from the values in the previous iteration. The second term describes the discrepancy between the observation and the prediction. The observation data is sparsely sampled from the fields and subjected to the Gaussian noise $\epsilon \sim \mathcal{N}(0, \mathsf{R})$, while the prediction is obtained by simulating the reservoir and extracting values at the observed locations through the operator $\mathcal{H}$.
The parameter-update scheme of the iterative ensemble Kalman method is written as
\begin{equation*}
  \label{eq:update}
  \begin{aligned}
    \bm{\omega}_j^{l+1} &= \bm{\omega}_j^l + \mathsf{K} \left(\mathsf{y}_j-\mathcal{H} [\bm{\omega}_j^l]\right), \\
    \text{with} \quad \mathsf{K} &= \mathsf{S}_{\omega}^l {(\mathsf{S}_d^l)}^\top \left[\mathsf{S}_d^l {(\mathsf{S}_d^l)}^\top + \mathsf{R}\right]^{-1},
  \end{aligned}
\end{equation*}
where $\mathsf{K}$ is the Kalman gain matrix, $\mathsf{S}_\omega^l$ and $\mathsf{S}_d^l$ are the square-root matrices with respect to the parameters and data, respectively, at iteration step $l$. The square-root matrices are calculated as
\begin{small}
\begin{gather*}
  \label{eq:square-root}
     \mathsf{S}_{\omega}^l = \frac{1}{\sqrt{N_e-1}}\left[\bm{\omega}_1^l-\bar{\bm{\omega}}^l, \ldots, \bm{\omega}_{N_e}^l-\bar{\bm{\omega}}^l\right], \\
     \mathsf{S}_d^l = \frac{1}{\sqrt{N_e-1}}\left[\mathcal{H}[\bm{\omega}_1^l] - \mathcal{H}[\bar{\bm{\omega}}^l], \ldots, \mathcal{H}[\bm{\omega}_{N_e}^l] - \mathcal{H}[\bar{\bm{\omega}}^l]\right], \\   
     \text{with} \quad \bar{\bm{\omega}}^l=\frac{1}{N_e} \sum_{j=1}^{N_e} \bm{\omega}_j^l,
\end{gather*}
\end{small}
where $N_e$ is the sample size.

In comparison to previous works using adjoint-based methods~\cite{li2001simultaneous,eydinov2009simultaneous}, the ensemble-based method has the advantages of being derivative-free and non-intrusive. It does not require the calculation for gradient of the cost function, and induces minimal modifications of the source code, making the implementation straightforward for a realistic reservoir. In this work, the forward propagation (Fig.~\ref{fig:schematic}b) is performed using the black-oil simulator, Open Porous Media (OPM) Flow~\cite{rasmussen2021open}, and the parameter update (Fig.~\ref{fig:schematic}c) is implemented with the DAFI code~\cite{strofer2021dafi}, while the complete inference procedure is carried out in a coupled manner.

\section{Results}\label{sec4}
We demonstrate the feasibility of the proposed method for inferring relative permeability curves through a series of test cases. The first two test cases, SPE 1 and SPE 3 benchmarks, are designed by the Society of Petroleum Engineers (SPE) as synthetic projects for evaluating and benchmarking various simulators or algorithms. These benchmark projects have become widely recognized within the petroleum engineering community as important tools for testing and improving the accuracy and efficiency of reservoir simulators. The third test case, Drogon, is a field-scale model that closely resembles certain features of a real field model, making it significantly more complex than the synthetic benchmarks described above.

For all test cases, we aim to infer the gas relative permeability curves from the sparse oil saturation data, while assuming the other relative permeability curves to be known. The inferred gas relative permeability curves are known as the ensemble mean curves after convergence, and are evaluated by comparing with the corresponding ground truths, which are provided in the OPM flow examples and assessed by the agreements with the commercial simulator ECLIPSE~\cite{rasmussen2021open}. In addition, the predicted saturation fields and well responses based on the inferred ensemble mean curves are compared to their equivalents obtained by using the true curves. Details of the case set-up and results are presented below.

\subsection{SPE 1 benchmark}\label{sec3-1}

\begin{figure}[!htb]
\centering
\includegraphics[height=0.29\textwidth]{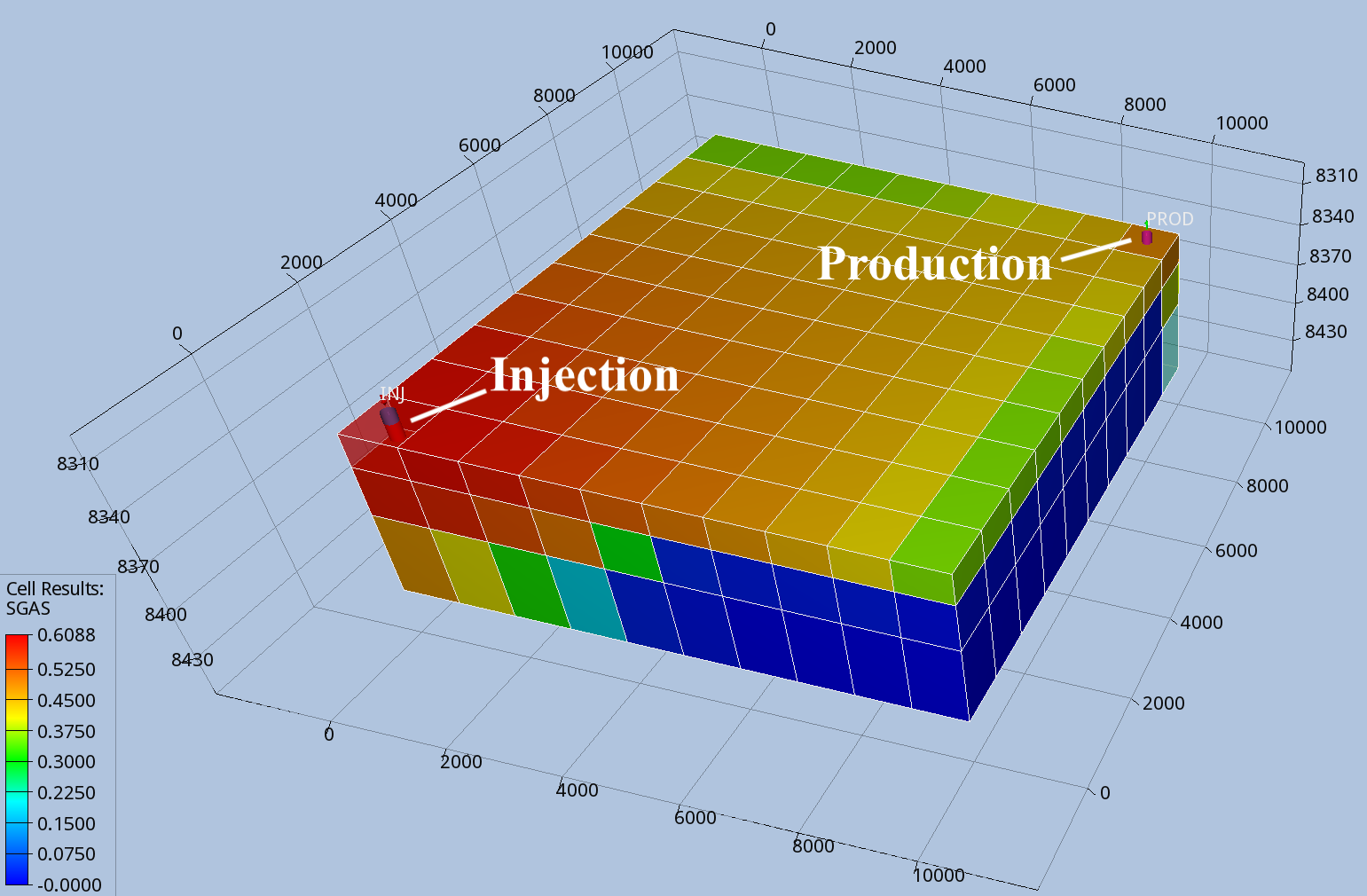}
  \caption{
  Computational domain and well locations of the SPE 1 benchmark. The domain length in $z$ direction is magnified 35 times for clarity. Two wells are located at the opposite corners for gas injection and oil production.
  }
  \label{fig:spe1-setup}
\end{figure}

The first case, SPE 1, is a benchmark project for three dimensional black-oil simulation. It has two suggested examples with different conditions: (1) constant saturation pressure and (2) varying saturation pressure for different gas saturations. Here we choose the second example to perform the inference 
since it was used to verify and validate the OPM Flow simulator~\cite{rasmussen2021open}.
The computational domain and well locations of this example are shown in Fig.~\ref{fig:spe1-setup}. Specifically, the reservoir has the domain size of $10000\ \mathrm{ft} \times 10000\ \mathrm{ft} \times 100\ \mathrm{ft}$, with thickness of 20 ft, 30 ft and 50 ft for each layer respectively. The domain is spatially discretized into $10 \times 10 \times 3$ cells for the numerical simulation. The first well (bottom left in Fig.~\ref{fig:spe1-setup}) injects gas at a rate of 100 MMscf/day from the top layer, while the other (top right) produces oil from the bottom layer. More details about SPE 1 benchmark can be found in the literature~\cite{rasmussen2021open,odeh1981comparison}. The reservoir is initially undersaturated with the gas saturation $S_\textrm{g} = 0$ over the whole domain. The reservoir is simulated over a period of ten years from Jan. 2015 to Dec. 2024. We use the oil saturations at 50 cells in late third year (Nov. 2017) as the observation data to infer the gas relative permeability curve. Note that no data from the last seven years is used for the inference; instead, the states of the reservoir over this period are predicted with the inferred gas relative permeability curve.

\begin{figure*}[ht]
\centering
\subfloat[Inferred curve]
{\includegraphics[height=0.4\textwidth]{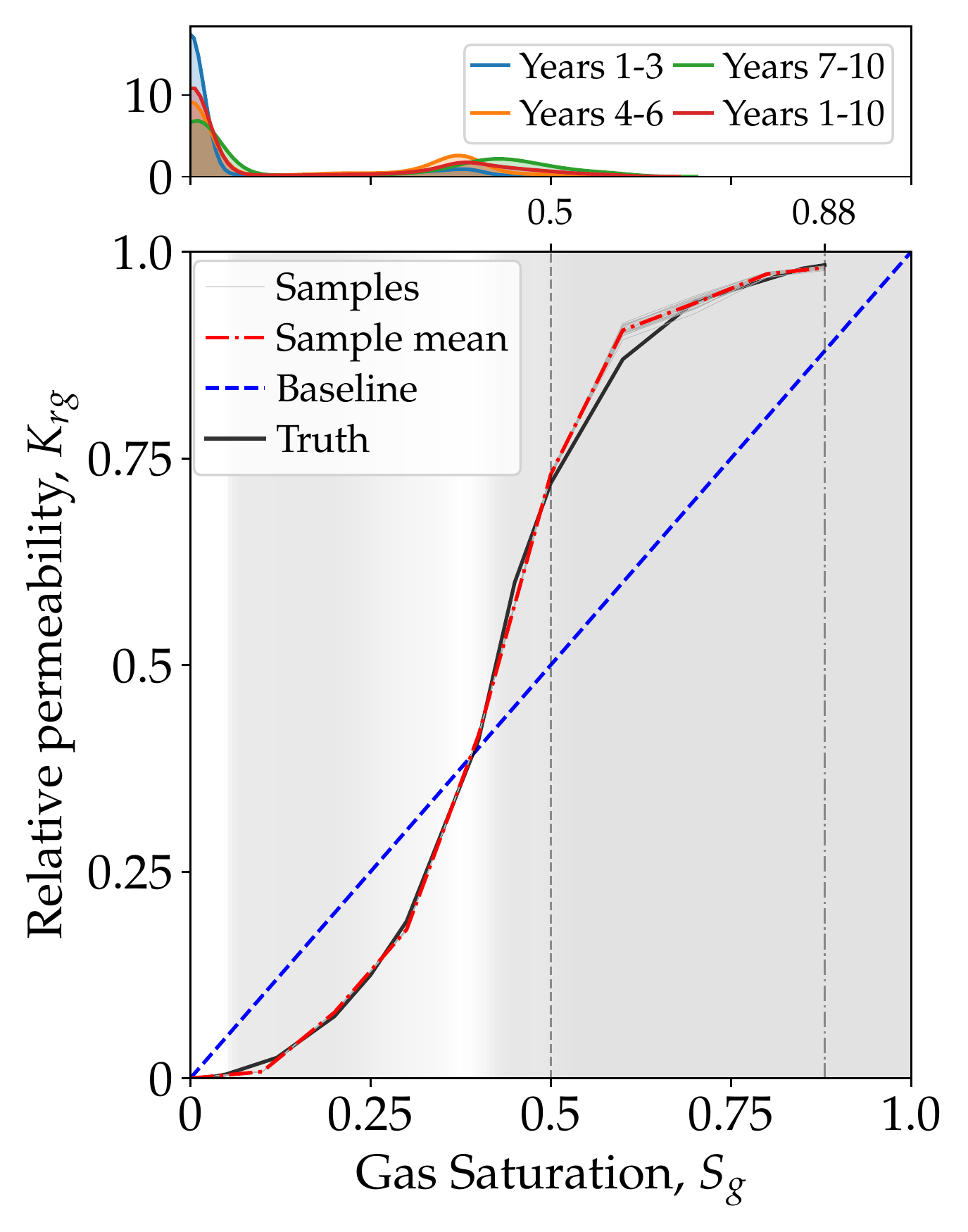}}
\hspace{1.0em}
\subfloat[Gas saturation field comparison]
{\includegraphics[height=0.4\textwidth]{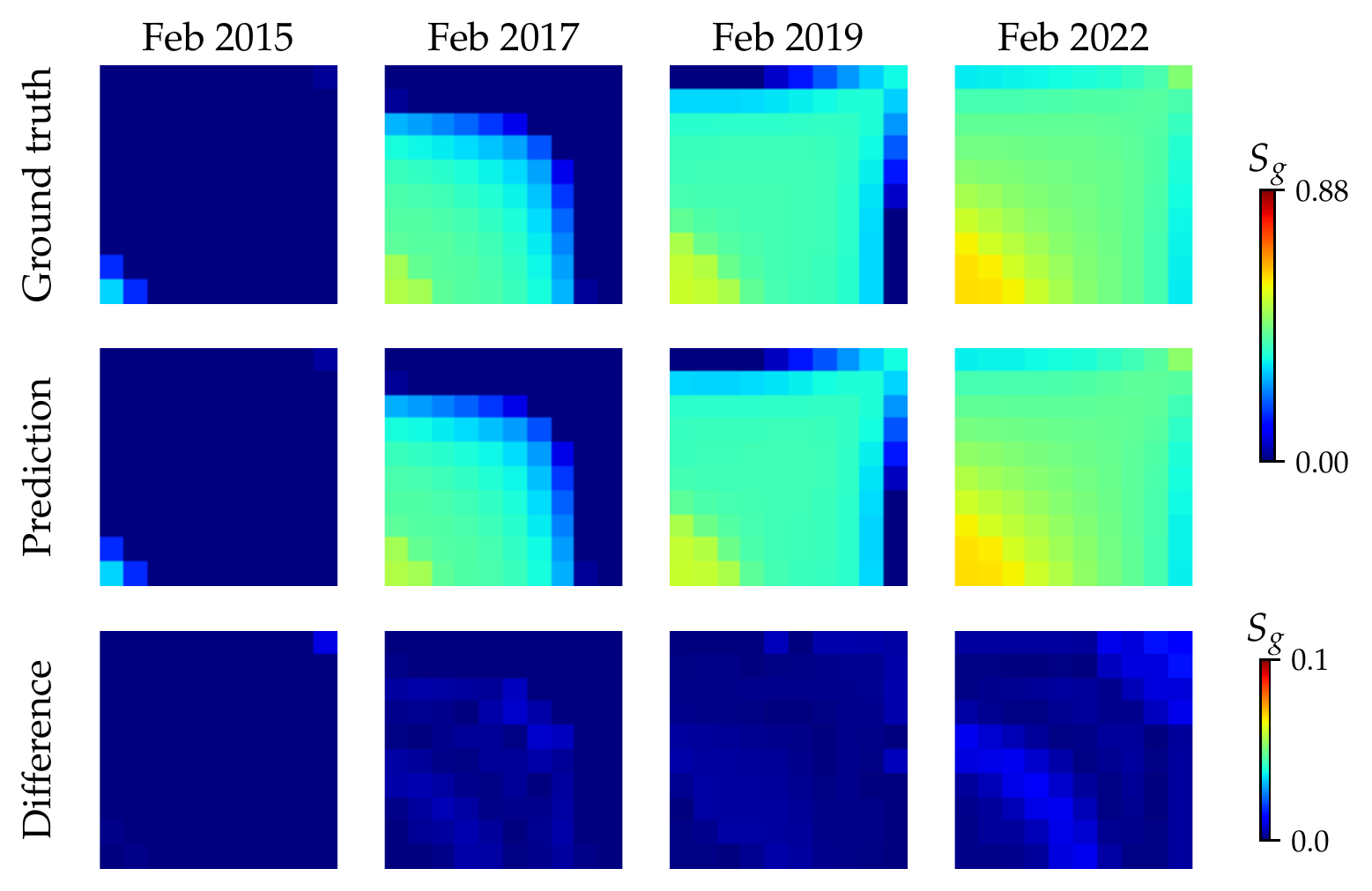}}
  \caption{Evaluation of the inferred relative permeability curve in the SPE 1 benchmark: (a) comparison of the inferred curve and the corresponding ground truth, along with the probability density functions (PDFs) of the gas saturations for different time periods, and (b) comparison of the predicted gas saturation fields at four distinct years based on the inferred and true relative permeability curves. The gradient background in (a) corresponds to the PDF for the first period, with the darker color representing a lower density of saturation data. The vertical line $S_\textrm{g} = 0.5$ represents the upper limit of the observation data, while the $S_\textrm{g} = 0.88$ indicates the connate water saturation is 0.12.
  \label{fig:spe1-learn}
  }
\end{figure*}

The inferred gas relative permeability curve is in good agreement with the ground truth, which is shown in Fig.~\ref{fig:spe1-learn}a. Despite the initial guess having a completely different form, all samples converge to the inferred curve, which is nearly identical to the ground truth. These curves end at $S_\textrm{g} = 0.88$ because of the existence of connate water, the saturation of which is $S_\textrm{wc} = 0.12$. Good performance in the inference of gas relative permeability curve can also be seen from the evolution of gas saturation in the first layer of the reservoir, which is shown in Fig.~\ref{fig:spe1-learn}b. We can observe a similar diffusion process from the injection corner to the opposite for both the prediction (middle row) and ground truth (top row). The predicted gas saturation fields are quite close to the ground truths for four different years, with only minor deviations (bottom row).

\begin{figure*}[!htb]
\centering
\includegraphics[width=0.88\textwidth]{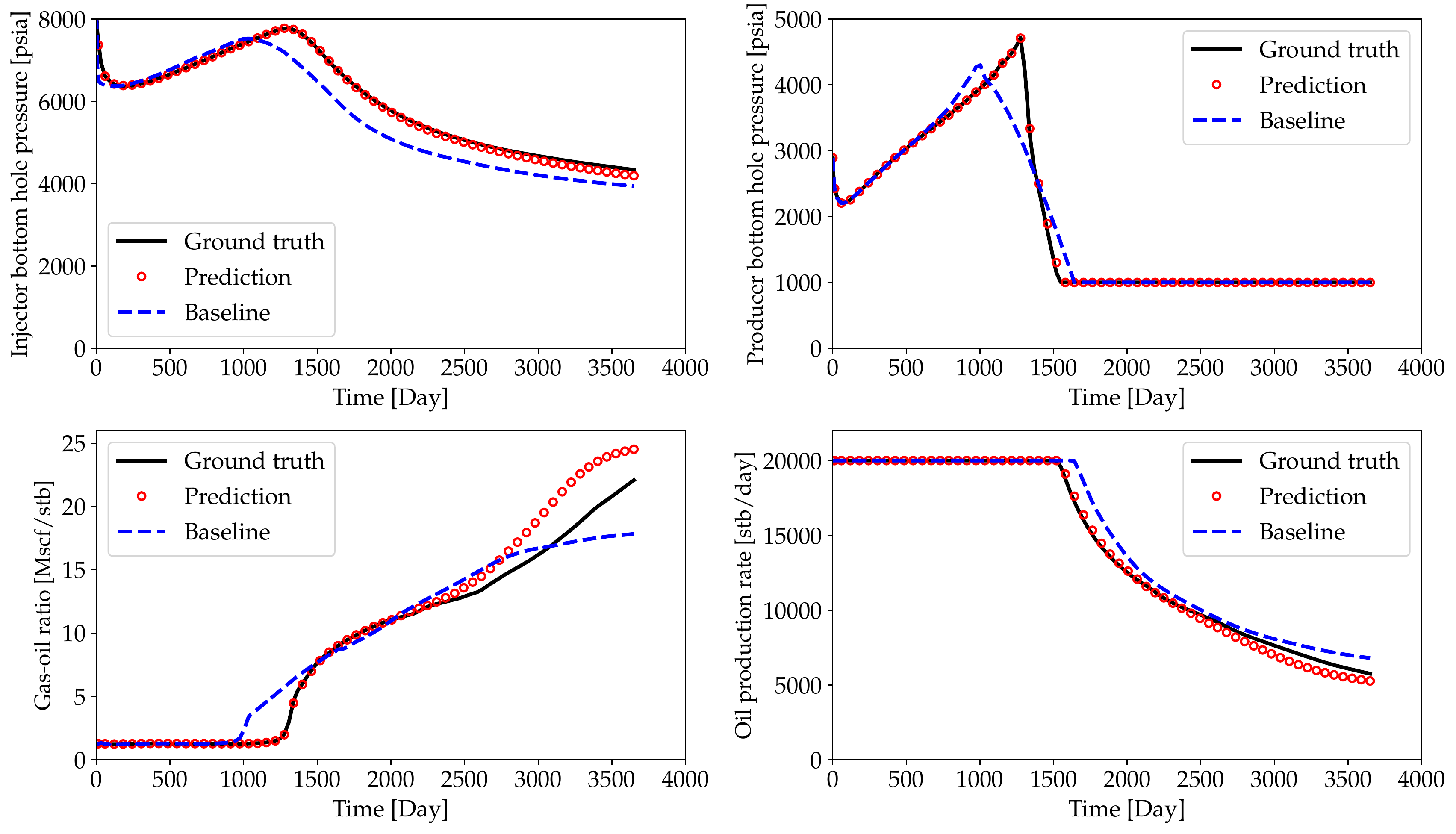}
  \caption{
  Comparison of the predicted well responses and the ground truths for the SPE 1 benchmark. Four quantities are selected for the comparison: (1) injection bottom hole pressure, (2) producer bottom hole pressure, (3) gas-oil ratio, and (4) oil production rate. The predictions and ground truths are obtained based on the inferred and true relative permeability curves, respectively, while the baselines are obtained using the initial guess of a straight line between (0, 0) and (1, 1).
  }
  \label{fig:spe1-compare}
\end{figure*}

The capability of the method is further demonstrated by the consistency for the important reservoir quantities (well responses), which is presented in Fig.~\ref{fig:spe1-compare}. It should be noted that none of these quantities are used as observation data in the inference of the relative permeability curve. Specifically, we evaluate the predictions for the time histories of four quantities by comparing them with the corresponding ground truths, i.e., (1) injection bottom hole pressure, (2) producer bottom hole pressure, (3) gas-oil ratio, and (4) oil production rate. The four predictions based on the inferred curve exhibit comparable trends to the ground truths. As a baseline for comparison, the initial guess, a line connecting (0, 0) and (1, 1), produces drastically different predictions. There are still discrepancies for the predicted gas-oil ratio and oil production rate after about 2500 days, which is due to the overestimate of gas relative permeability in the saturation interval [0.5, 0.7] shown in Fig.~\ref{fig:spe1-learn}a. The overestimate implies a greater ability to transmit gas in the presence of oil, leading to a simulation result with larger gas saturation and smaller oil saturation. Such an overestimate is consistent with that in the gas-oil ratio and the underestimate in the oil production rate (bottom row, Fig.~\ref{fig:spe1-compare}).

The discrepancy between the inferred and true gas relative permeability curves can be explained by the data distribution shown in upper panel of Fig.~\ref{fig:spe1-learn}a. We plot the probability density functions (PDFs) of gas saturations over the entire reservoir for three consecutive time periods (years 1--3, 4--6, and 7--10), while only the observation data in the first period is used for the inference. The data distribution of gas saturation in the first period is also presented by the gradient background in the bottom panel, where the darker background indicates the interval with a lower density of data. The curve is well estimated in the interval [0, 0.5] with a lighter background, while showing tiny deviations in the interval [0.5, 0.7] due to the paucity of observation data. Because of a lack of observation data, it is impossible to infer the curve within the interval [0.5, 0.88]. However, it happens to fit the ground truth because the portion of the curve that can be inferred covers a bulk of the relative permeability range from 0 to roughly 0.75, making the remaining piece of the curve well approximated with the constraints of monotonicity and boundedness.

Additionally, we employ production data as the observation to infer the relative permeability curve in SPE 1, as has been the common practice in many previous works. The results show clear ill-posedness for the inference due to the use of production data only.
More details can be found in Appendix~\ref{secA1}.

\subsection{SPE 3 benchmark}

\begin{figure}[!htb]
\centering
\includegraphics[height=0.26\textwidth]{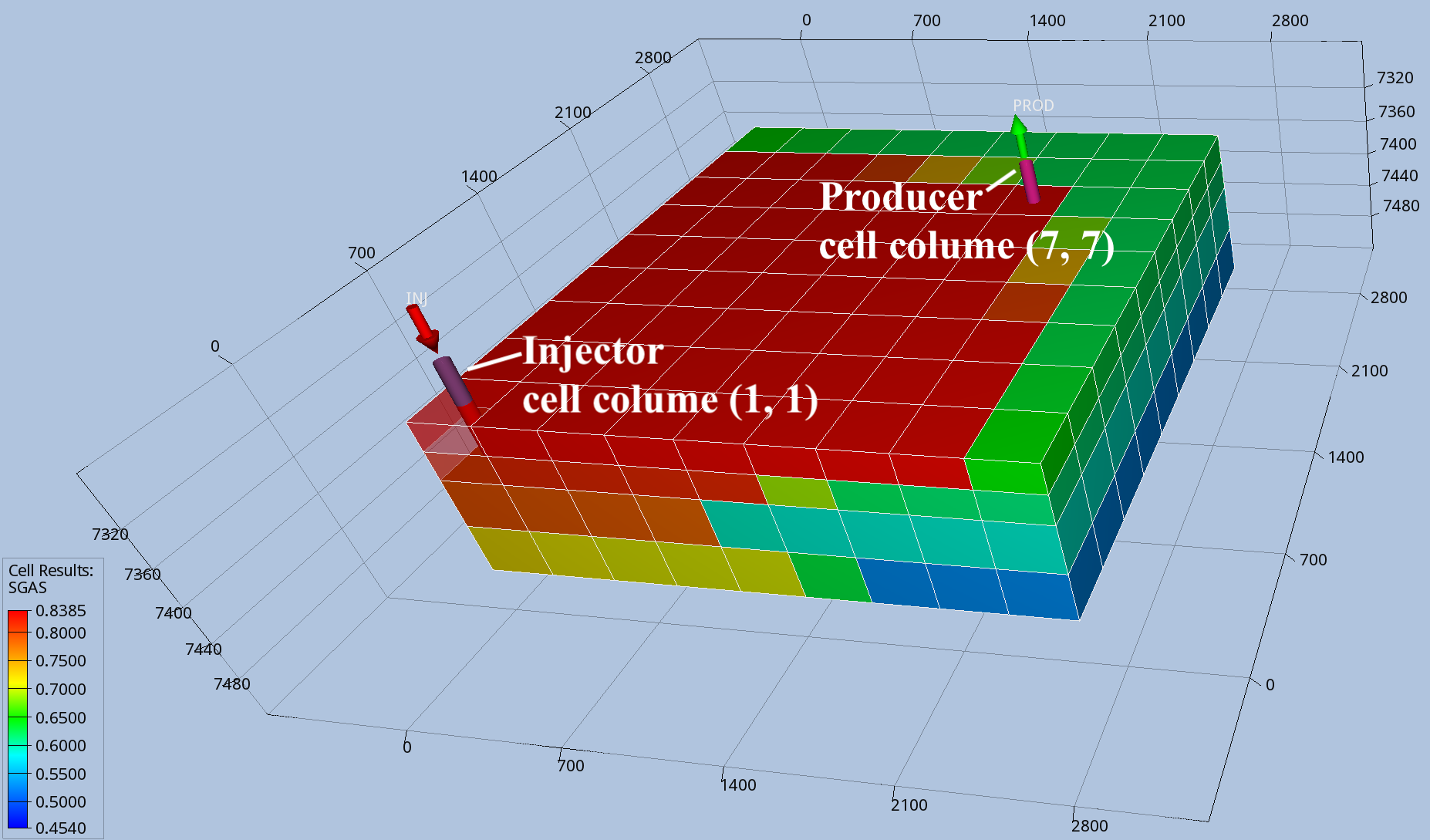}
  \caption{
  Computational domain and well locations of the SPE 3 benchmark. The domain length in $z$ direction is magnified 5 times for clarity. The injector perforates the top two layers and producer perforates the bottom two layers, and they are located in the cell columns of (1, 1) and (7, 7), respectively.
  }
  \label{fig:spe3-setup}
\end{figure}

The second case, SPE 3, is a benchmark introduced for studying gas cycling of retrograde condensate reservoirs. Gas condensate reservoirs refer to natural gas systems that exist in reservoirs with initial temperatures ranging from the critical temperature to the cricondentherm. In this case, the initial reservoir pressure is above the dew-point pressure, resulting in a single-phase gas system within the reservoir. The computational domain and well locations are shown in Fig.~\ref{fig:spe3-setup}.
The grid has a dimension of $9 \times 9 \times 4$ and each cell measures $293.3\ \mathrm{ft} \times 293.3\ \mathrm{ft}$ in the horizontal direction. The thickness of each layer is $30\ \mathrm{ft}$ for the first two layers and $50\ \mathrm{ft}$ for the last two layers.
The gas injector is positioned in the top corner of the cell grid at cell column (1, 1) and perforates the top two layers of cells, while the producer is located in the bottom corner of the cell grid at cell column (7, 7) and perforates the bottom two layers of cells. The reservoir is initially filled with gas and water with almost no oil, and the connate water saturation $S_\textrm{wc} = 0.16$ is assumed to be known here.
The gas reservoir is simulated over a 14-year period from Jan. 2016 to Dec. 2029. The gas injection rate in the reservoir is 5700 Mscf/day for the first four years, followed by a reduction to 3700 Mscf/day for the next five years, and finally, no injection during the last five years. More details of the case set-up can be found in the literature~\cite{rasmussen2021open,kenyon1987third}. We use the oil saturations at 50 cells in the late second year as the observation data to infer the gas relative permeability curve, while predicting the states of the reservoir over the next 12 years. 

\begin{figure*}[ht]
\centering
\subfloat[Inferred curve]
{\includegraphics[height=0.4\textwidth]{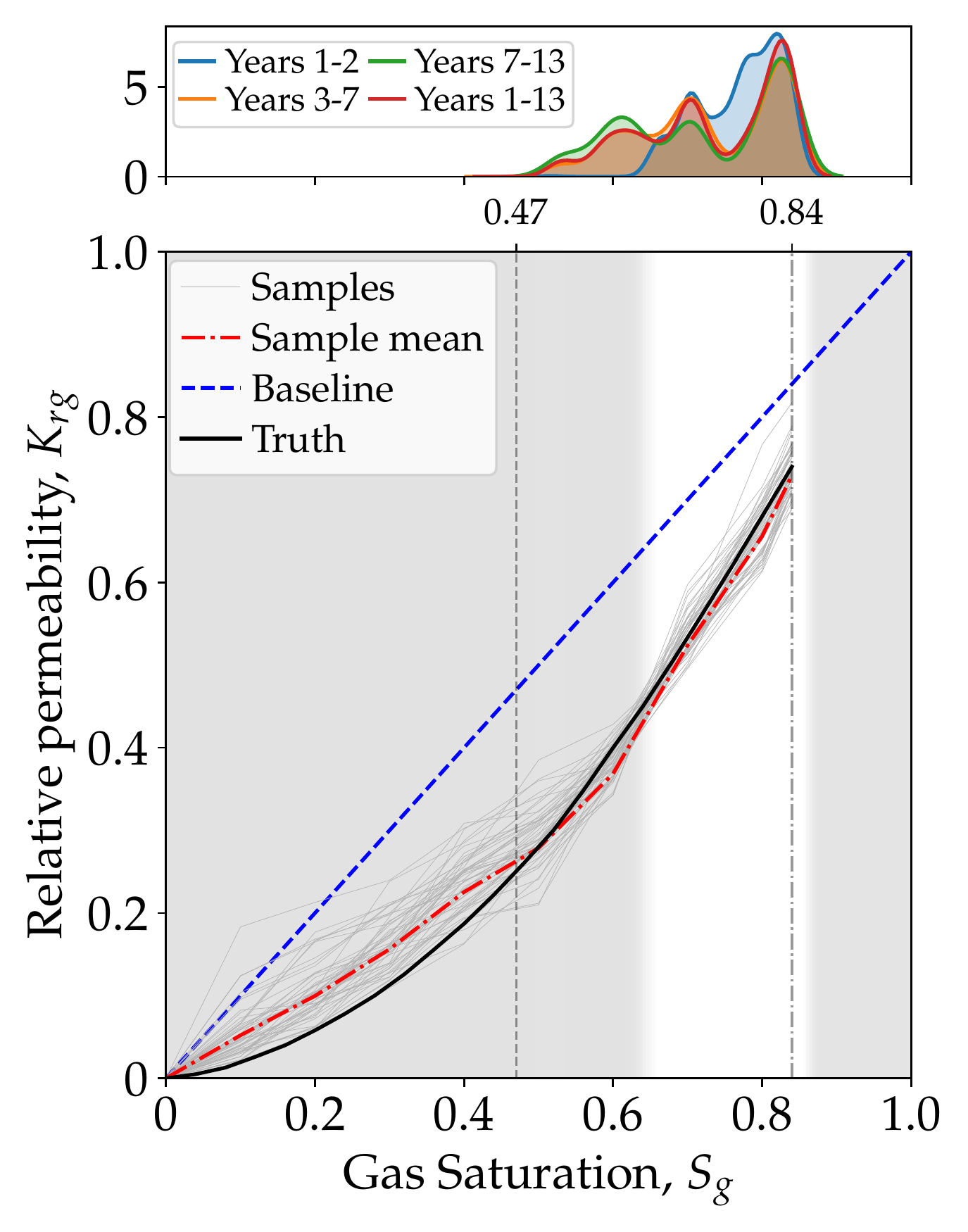}}
\hspace{1.0em}
\subfloat[Gas saturation field comparison]
{\includegraphics[height=0.4\textwidth]{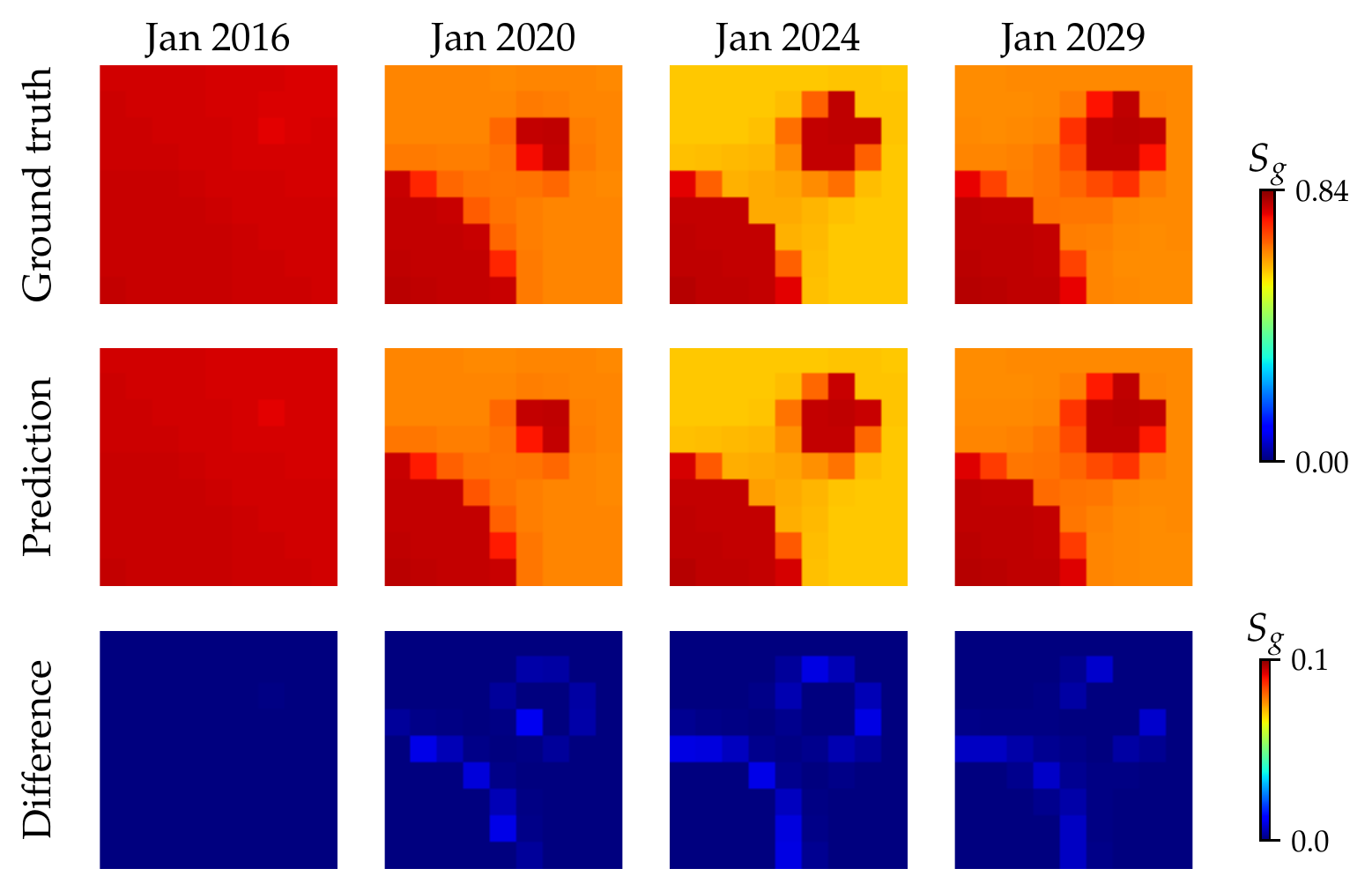}}
  \caption{Evaluation of the inferred relative permeability curve in the SPE 3 benchmark: (a) comparison of the inferred gas relative permeability curve and the ground truth, along with the PDFs of the gas saturation for different time periods, and (b) comparison of the predicted gas saturation fields at four distinct years based on the inferred and true curves. The gradient background in (a) corresponds to the PDF for the first two years, with the darker color representing a lower density of the saturation data. The vertical line $S_\textrm{g} = 0.47$ represents the lower limit of the observation data, while the $S_\textrm{g} = 0.84$ indicates the connate water saturation is 0.16.
  \label{fig:spe3-learn}
  }
\end{figure*}

\begin{figure*}[!htb]
\centering
\includegraphics[width=0.88\textwidth]{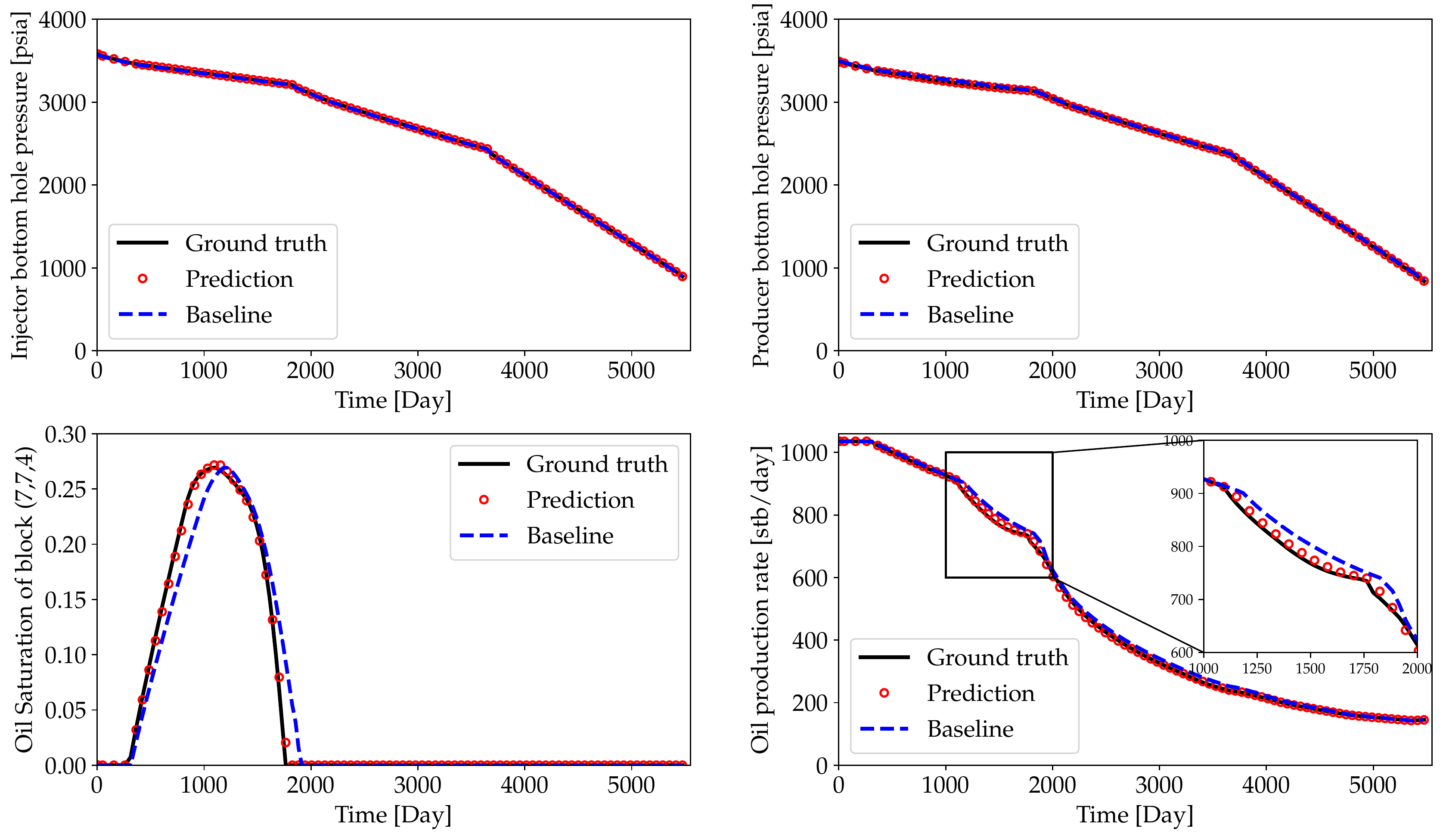}
  \caption{
  Comparison of the predicted well responses and the ground truths for the SPE 3 benchmark. Four quantities are selected for the comparison: (1) injector bottom hole pressure, (2) producer bottom hole pressure, (3) oil saturation of cell (7, 7, 4), and (4) oil production rate. The predictions and ground truths are obtained from the inferred and true curves, while the baselines are provided by the initial guess of a straight line between (0, 0) and (1, 1).
  }
  \label{fig:spe3-compare}
\end{figure*}

The inferred gas relative permeability curve is close to the ground truth, which is shown in Fig.~\ref{fig:spe3-learn}a. The sample curves diverge from the initial guess and converge to the inferred relative permeability curve after a few iterations. The performance is further presented by the minor differences (bottom row) between the predicted saturations (middle row) in the third layer of the reservoir and the corresponding ground truths (top row) for four different years, as is shown in Fig.~\ref{fig:spe3-learn}b. Similarly, the time series of the predicted well responses are nearly identical to those obtained from the true curve, as illustrated in Fig.~\ref{fig:spe3-compare}, which demonstrates the validity of this method. Nonetheless, the baselines of the well responses obtained from the initial guess also present similarity to the ground truths, indicating that the well responses are insensitive to the variation of relative permeability curve in this case. Such a result corresponds to our findings in the Appendix~\ref{secA1} that incorporating saturation as observation data can help lessen the ill-posedness for inferring the relative permeability curves.

The discrepancy between the inferred relative permeability curve and the ground truth can be explained by the data distributions in the upper panel of Fig.~\ref{fig:spe3-learn}a. The gas saturation data for the first two years are spread within the interval [0.47, 0.84], meaning that the relative permeability curve cannot be inferred within the interval [0, 0.47] due to the lack of observation data. This is also reflected by the clear variance of the sample curves in the same range. Despite the existence of saturation data in the range [0.47, 0.65], the probability density is too low to estimate the curve accurately, as illustrated by the dark background. The majority of the saturation data is distributed within the interval [0.65, 0.84], allowing for a reliable estimation within this range.

\subsection{Drogon case}

\begin{figure}[!htb]
\centering
\includegraphics[width=0.47\textwidth]{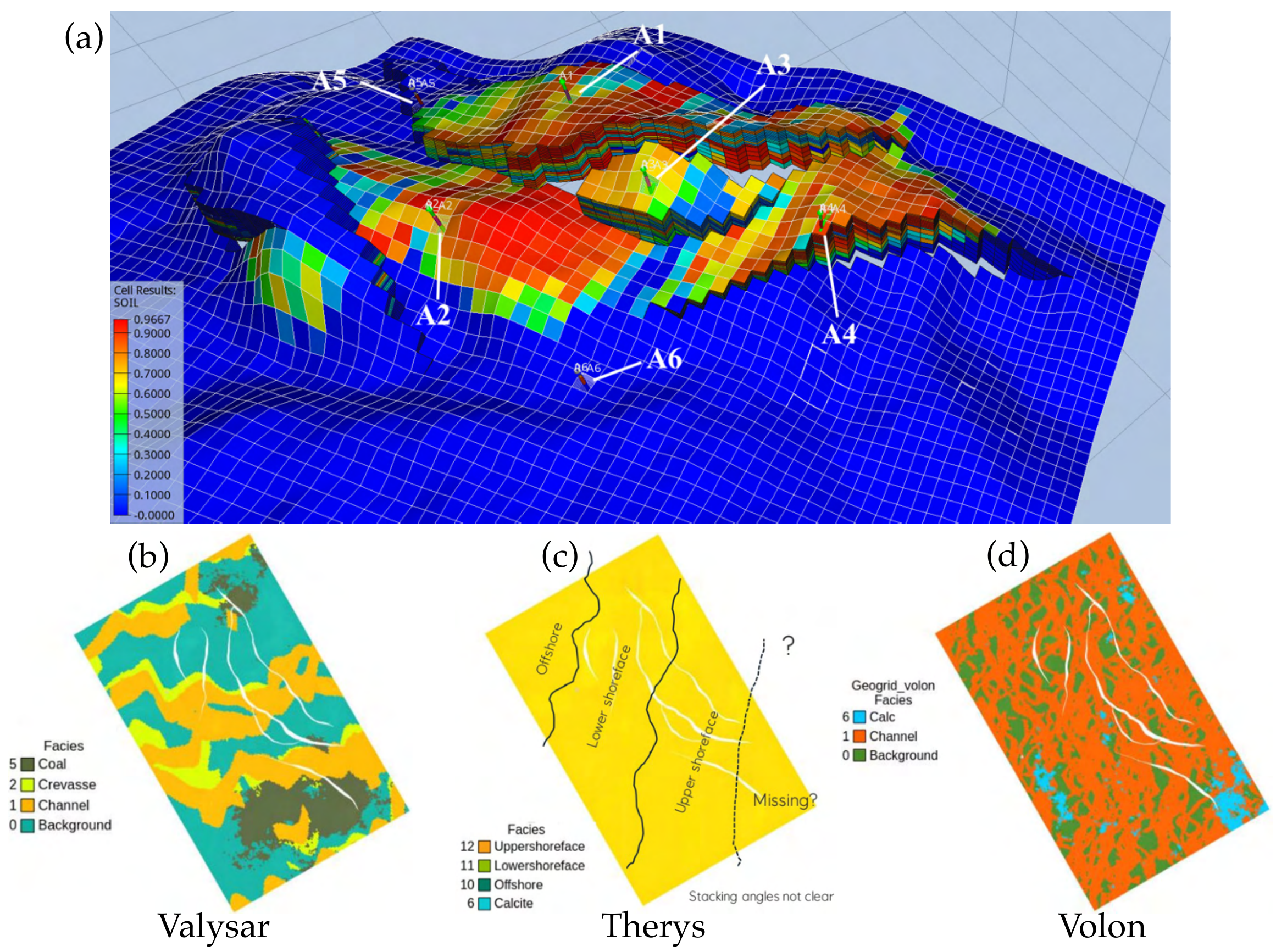}
  \caption{
  Graphical description of the Drogon case: (a) computational domain and well locations, and (b), (c), and (d) for three designed formations therein. The domain length in $z$ direction is magnified 10 times for clarity. A1, A2, A3 and A4 are oil producers, whereas A5 and A6 inject water.
  }
  \label{fig:drogon-setup}
\end{figure}

The proposed method is further evaluated in a more complex case called Drogon, which is shown in Fig.~\ref{fig:drogon-setup}a. It is the successor to the Reek Field~\cite{chang2022ensemble}, both of which are developed by Equinor. According to the conceptual description~\cite{equinor2020drogon}, the Drogon reservoir is located in the Volantis Group, which comprises three formations: Valysar, Therys, and Volon, as depicted in Figs.~\ref{fig:drogon-setup}b-d, respectively. Specifically, the Valysar formation is a fluvial system with channel bodies and its lower boundary is dominated by a continuous coal; the Therys formation is a shoreline system grading from shoreface facies to offshore facies, with calcite cemented strings; and the Volon formation is a braided fluvial system with the presence of calcites intervals. The Drogon reservoir has an approximate domain size of $12000\ \mathrm{ft} \times 15000\ \mathrm{ft} \times 450\ \mathrm{ft}$ and is spatially discretized into $46\times73\times31$ cells, with 70972 active cells. The complexity of the reservoir is also reflected by the presence of 12 saturation table regions (SATNUMs) with different relative permeability curves for each, as opposed to the above synthetic cases (SPE 1 and SPE 3) where the relative permeability curves keep identical over the entire fields. Two wells (A5 and A6) near the oil-water contact inject water with a maximum injection rate of 8000 SM$^3$/day, while the other four wells (A1--A4) throughout the oil-containing area produce oil with a maximum production rate of 4000 SM$^3$/day.

The Drogon reservoir is simulated over a period of two and a half years from Jan. 2018 to Jul. 2020. The injectors, A5 and A6, are opened on May 2018 and Nov. 2018, respectively, and are briefly shut down for three times during this period; the producers, A1--A4, are opened in sequence. During the simulation, the majority of saturation changes are confined to a single region (SATNUM 1), while the remaining regions (SATNUMs 2-12) exhibit little changes. Here we intend to infer the gas relative permeability curve in SATNUM 1, given that others curves are already known. We use the oil saturations at 85 cells in the last month (Jul. 2020) as the observation data to perform the inference, with the cells picked to have the largest saturation changes.

\begin{figure*}[ht]
\centering
\subfloat[Inferred curve]
{\includegraphics[height=0.39\textwidth]{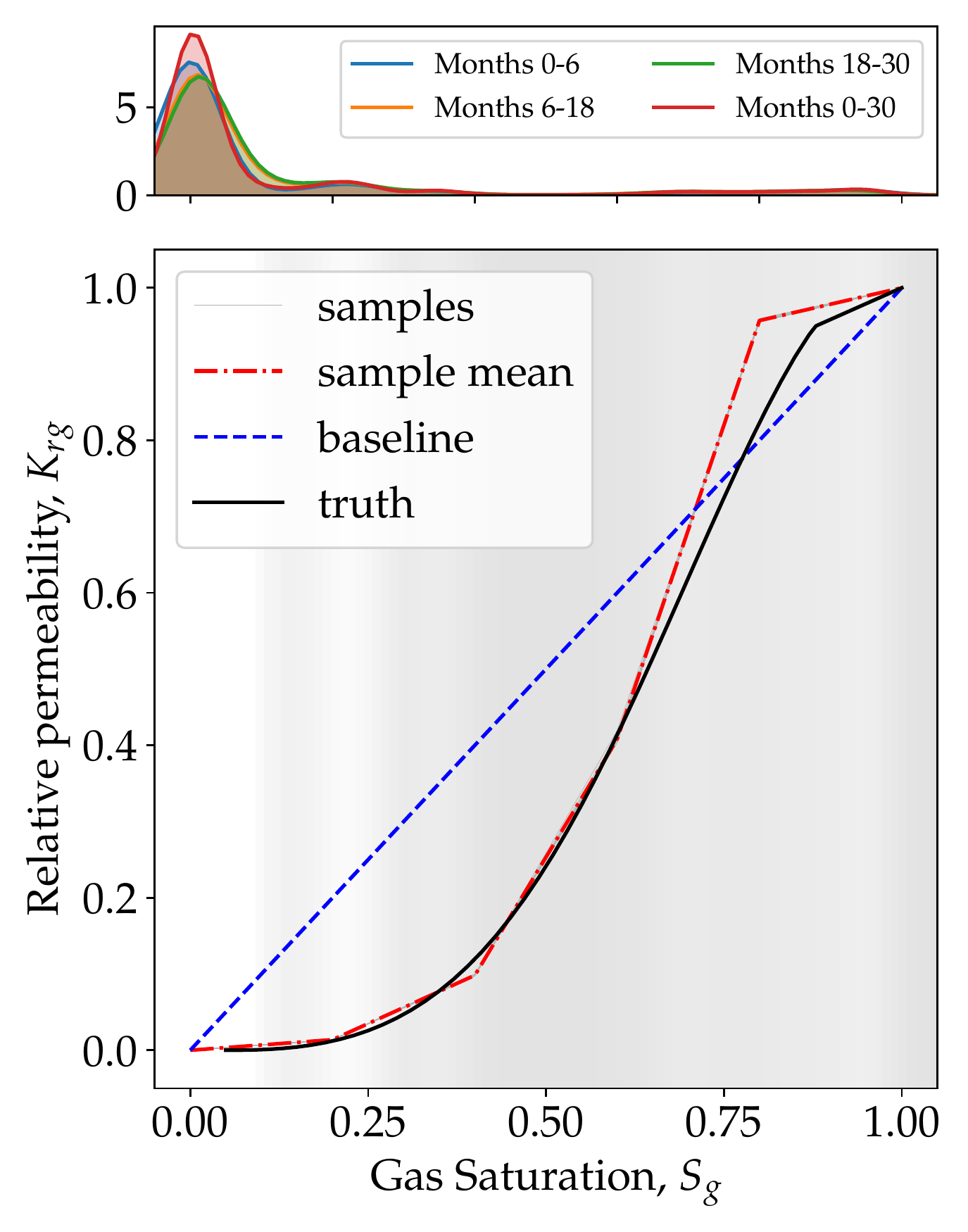}}
\hspace{1.0em}
\subfloat[Gas saturation field comparison]
{\includegraphics[height=0.39\textwidth]{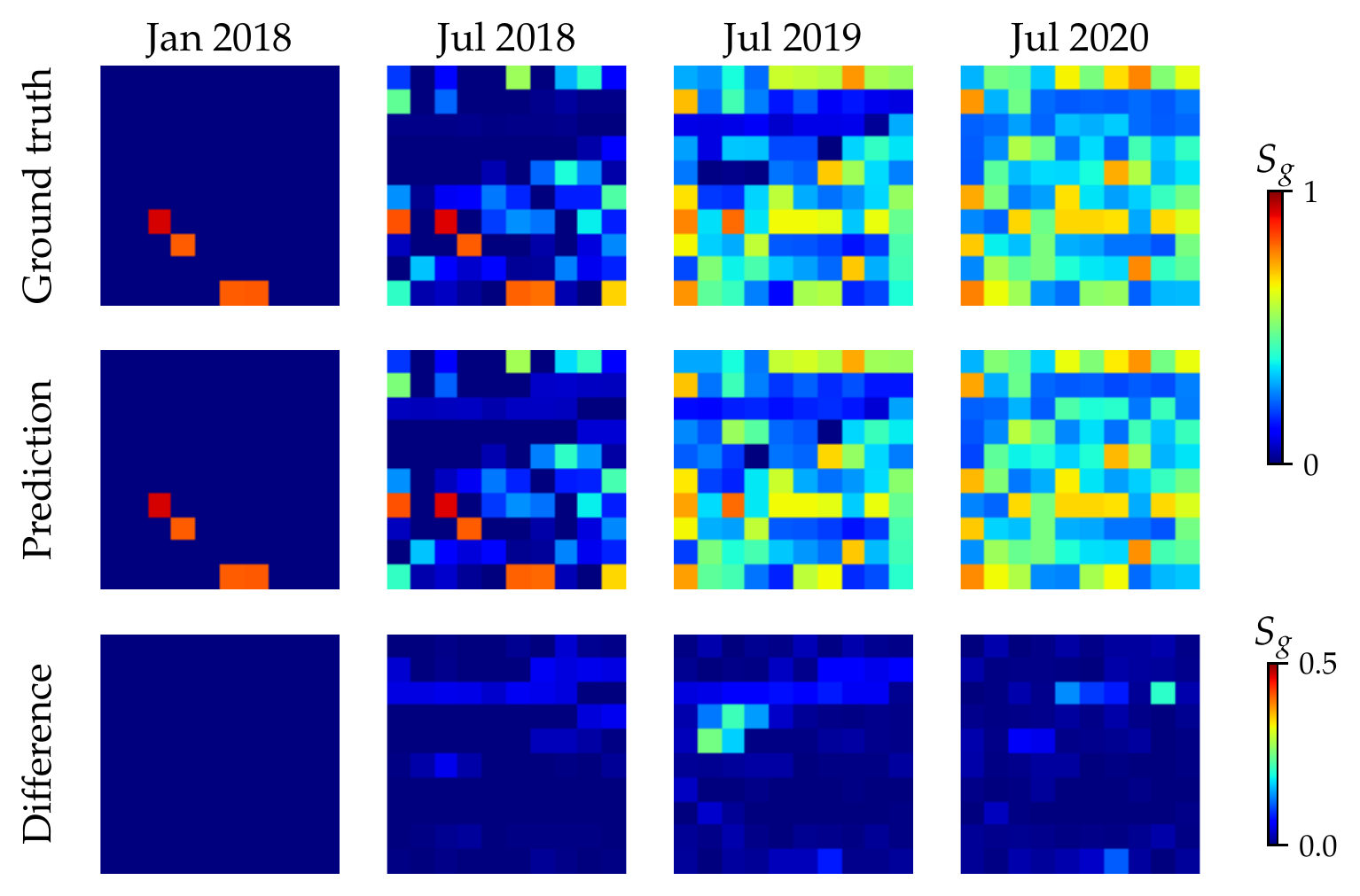}}
  \caption{Evaluation of the inferred relative permeability curve in the Drogon case: (a) comparison of the inferred curve and the ground truth, along with the PDFs of the gas saturation data for different time periods, and (b) comparison of the predicted gas saturation fields at four distinct months based on the inferred and true curves. The gradient background in (a) corresponds to the PDF for the entire time period, with the darker color representing a lower density of data.
  \label{fig:drogon-learn}
  }
\end{figure*}

The inferred gas relative permeability curve has a comparable form to the corresponding ground truth despite some deviation for high saturations, which is illustrated in Fig.~\ref{fig:drogon-learn}a. To be specific, the relative permeability curve is accurately estimated within the saturation interval [0, 0.6], but overestimated within the interval [0.6, 1]. The performance is reasonable given that the gas saturation values in most of the cells are less than 0.6, as shown in the PDF plot in the upper panel, where the data distribution is skewed with the majority of values concentrated near zero. The inference performance is further explained by the gradient background: The relative permeability curve is correctly inferred throughout the interval with a light background, but deviates within the interval having a darker background, for which much fewer observation data are available.

\begin{figure*}[!htb]
\centering
\includegraphics[width=0.88\textwidth]{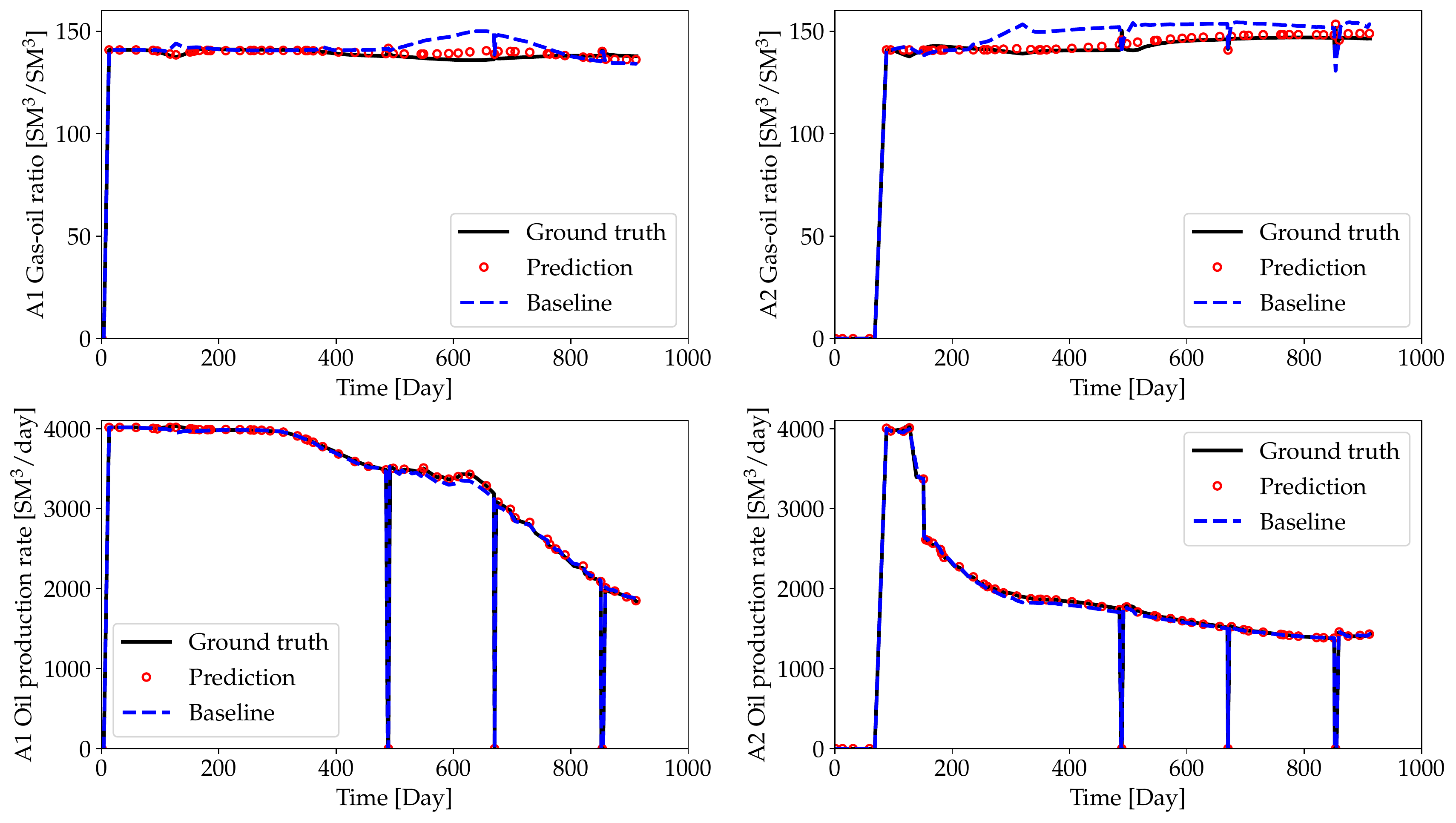}
  \caption{
  Comparison of the predicted well responses and the ground truths for the Drogon case. Four quantities are selected for the comparison in two injectors: (1) A1 gas-oil ratio, (2) A2 gas-oil ratio, (3) A1 oil production rate, and (4) A2 oil production rate. The predictions and ground truths are obtained from the inferred and true curves, while the baselines are provided by the initial guess of a straight line between (0, 0) and (1, 1).
  }
  \label{fig:drogon-compare}
\end{figure*}

Despite some discrepancy between the inferred and true curves, the predicted reservoir states are comparable to the corresponding ground truths, which is demonstrated in Figs.~\ref{fig:drogon-learn}b and~\ref{fig:drogon-compare}. In Fig.~\ref{fig:drogon-learn}b, the predicted gas saturations (middle row) at 225 ($15 \times 15$) cells for four separate months are quite close to the ground truths (top row) obtained from the true curve, despite differences (bottom row) at very few locations. Note that the 225 cells are not confined to a certain layer, but are instead selected to have the most pronounced saturation changes among all 70972 cells. Similarly, as shown in Fig.~\ref{fig:drogon-compare}, the predicted well responses exhibit comparable trends to those of the corresponding ground truths but differs from the baselines, particularly for the gas-oil ratio (top row). Specifically, we compare the predicted gas-oil ratio and oil production rate with the ground truths and baselines for the producers A1 and A2. The predictions are in good agreement with the ground truths, while the baselines display overestimates over a long time period. The overestimate is a result of the greater gas transmission ability implied in the initial guess, which has larger relative permeabilities in the saturation interval [0, 0.8] than those of the true curve. The oil production rate (bottom row) again demonstrates its insensitivity to the gas relative permeability curve, necessitating the use of saturation data to regularize the ill-posedness that can be induced by history matching the production data alone.

\section{Conclusion}\label{sec5}
This paper demonstrates an ensemble-based framework for inferring the relative permeability curves from sparse saturation data. In regards to representing relative permeability curves, this suggested paradigm differs from the previous works. The inherent monotonicity and boundedness of the representation eliminates the need to impose extra constraints during the optimization procedure. In addition, we try using sparse saturation data as the observation, which differs from commonly used production data and can be obtained via seismic monitoring. We find that incorporating the saturation data helps lessen the ill-posedness of inferring the curves when compared to the use of production data only.

The capability of the framework is proved on two synthetic benchmarks and a field-scale case with certain real-field features. All of these tests demonstrate that the proposed method is capable of inferring the relative permeability curves, and the third case demonstrates its potential for use in real fields. The relative permeability curves can be accurately estimated for saturations with available data, while the embedded constraints ensure extrapolating to other saturations appropriately.

\backmatter

\bmhead{Acknowledgments}

The authors thank Xin-Lei Zhang (Chinese Academy of Sciences), Hongsheng Wang (University of Texas at Austin), and Tor Harald Sandve (NORCE Norwegian Research Centre AS) for their valuable discussions and suggestions.

\section*{Declarations}

\subsection*{Funding}
This work was financially supported by the University Coalition for Fossil Energy Research (UCFER) Program under the U.S. Department of Energy’s National Energy Technology Laboratory through the Award No. DE-FE0026825 and SubAward No. S000038-USDOE.

\subsection*{Conflict of interest}
The authors have no relevant financial or non-financial competing interests.

\subsection*{Code availability}
The code is available on GitHub~\cite{zhou2023krcurve-git}, which can be used by the readers for reproducing the results and further development.

\subsection*{Author contribution statement}
JM, CC, and HX supervised the project. XZ, HW, and JM performed the research. XZ and HW wrote the manuscript. JM, CC, and HX provided insightful feedback and helped shape the research, analysis, and manuscript.

\begin{appendices}

\section{Inference with production data}\label{secA1}

In this paper, we conduct two more numerical experiments with SPE 1 benchmark to highlight the ill-posedness induced by using production data only, as is the norm in many research of this nature.

\begin{figure*}[ht]
\centering
\subfloat[Inferred curve]
{\includegraphics[height=0.36\textwidth]{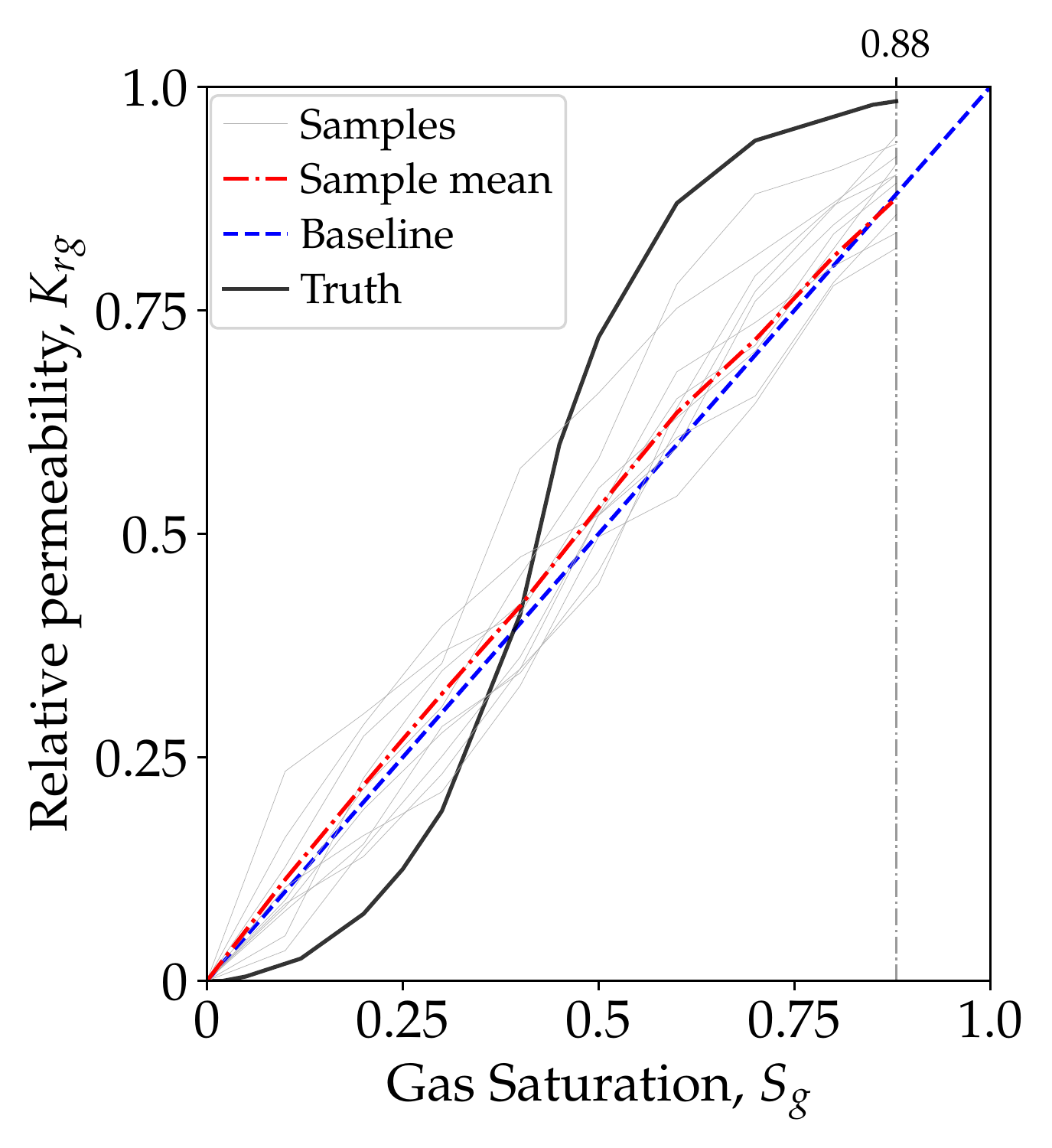}}
\hspace{0.5em}
\subfloat[Gas saturation field comparison]
{\includegraphics[height=0.36\textwidth]{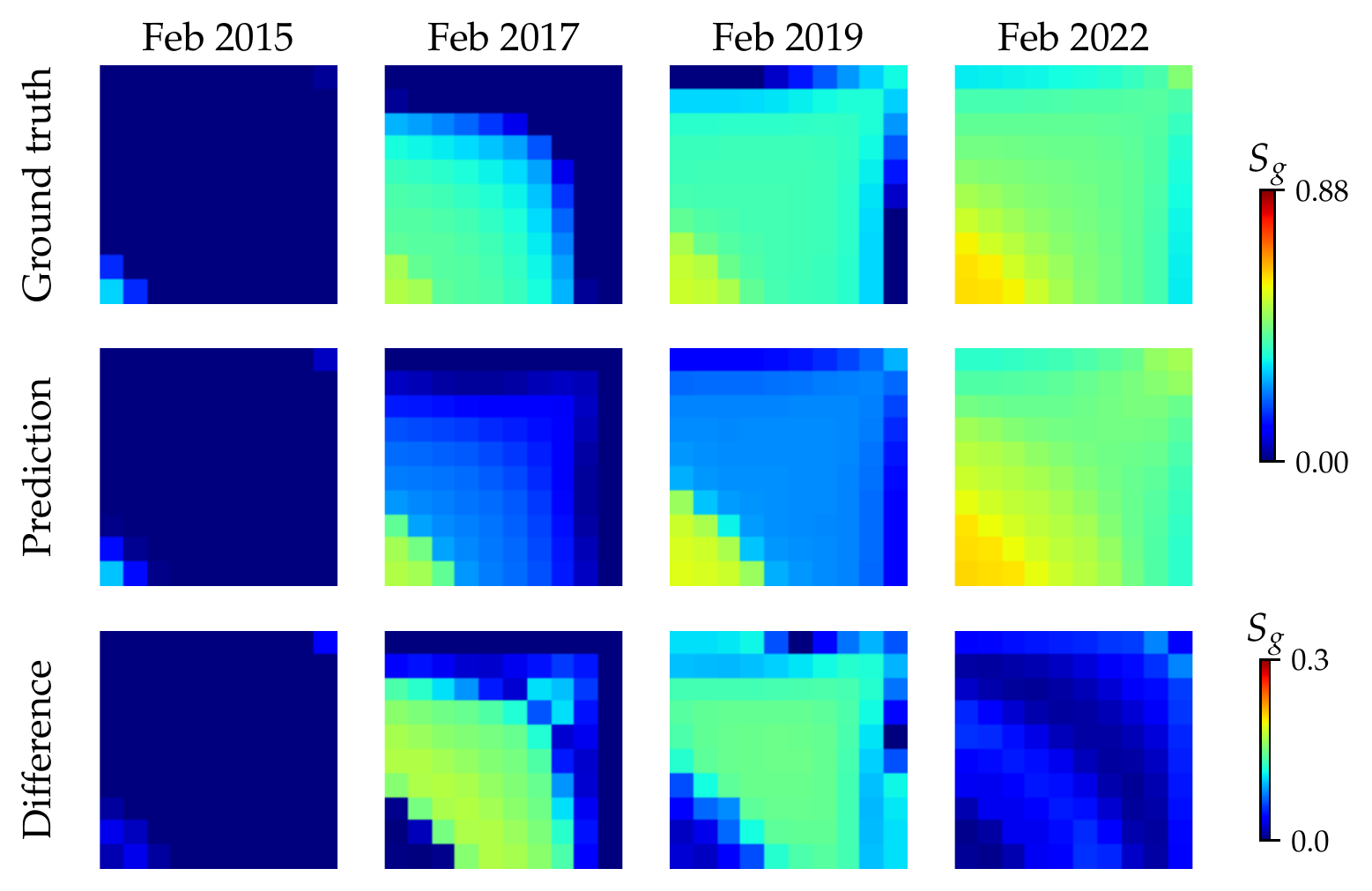}}\\
\subfloat[Well response comparison]
{\includegraphics[width=0.88\textwidth]{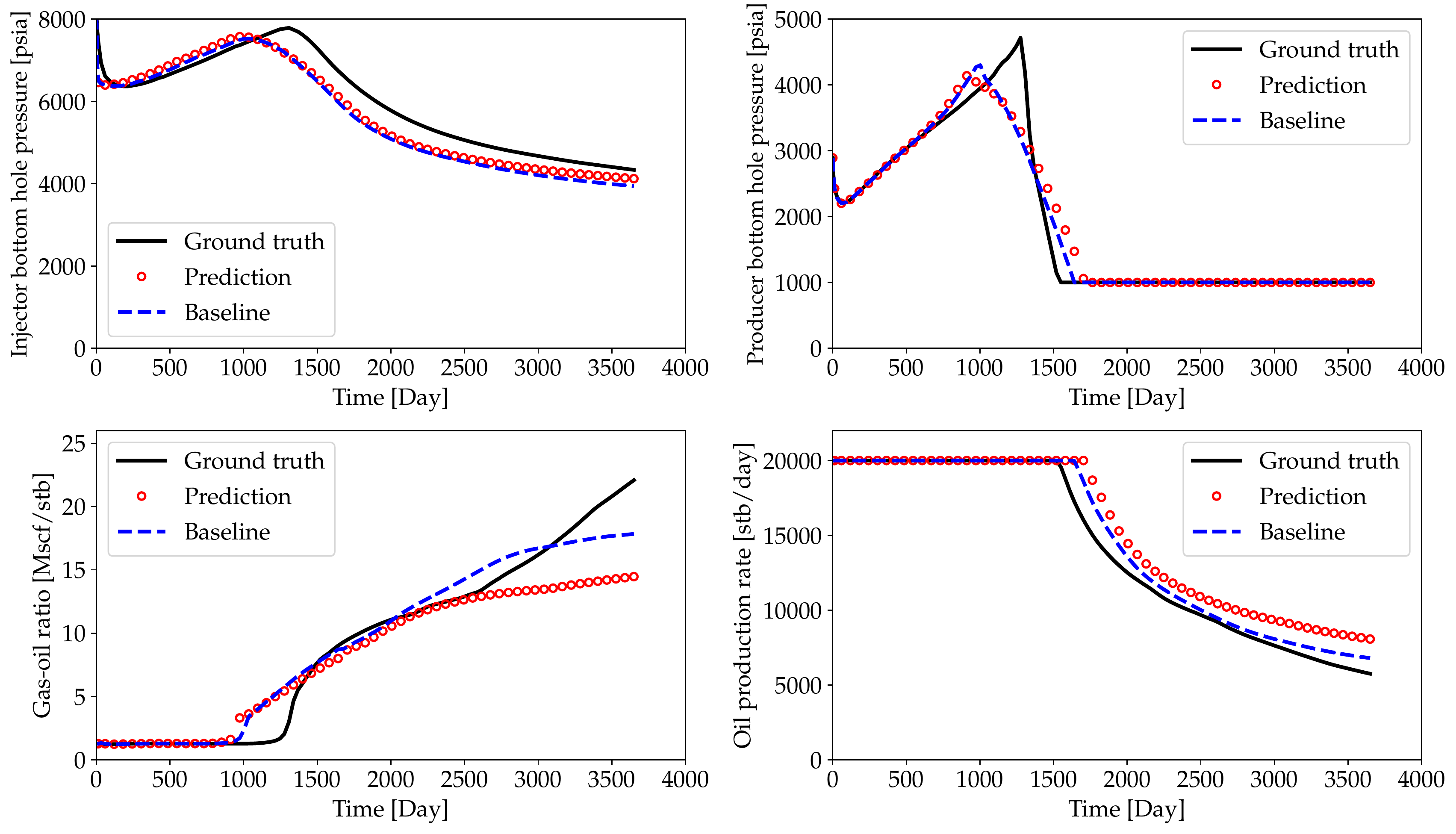}}
  \caption{Evaluation of the inferred relative permeability curve in SPE 1 benchmark: (a) comparison of the inferred curve with the ground truth, (b) comparison of the simulated gas saturation fields at four distinct years based on the inferred and true curves, and (c) comparison of the time series of well responses over 10 years based on the baseline, inferred and true curves. Here, the oil production rate time series for the first three years are used as the observation data.
  \label{fig:oildata-35}
  }
\end{figure*}

In both experiments, the inferred gas relative permeability curve diverges significantly from the ground truth. The first experiment is designed for a fair comparison with that described in Sec.~\ref{sec3-1}. We infer the relative permeability curve using the oil production rate in the first three years (Jan. 2015 to Dec. 2017) as the observation data. The inference result is shown in Fig.~\ref{fig:oildata-35}a. The sample curves do not converge after iterations and their mean is nearly equal to the initial guess.
Consequently, the predicted gas saturation fields and well responses are significantly different from the corresponding ground truths, as shown in Figs.~\ref{fig:oildata-35}b and c, respectively. The performance is expected because the initial guess can yield the same oil production rate for the first three years, rendering the inference by definition ill-posed.

\begin{figure*}[ht]
\centering
\subfloat[Inferred curve]
{\includegraphics[height=0.36\textwidth]{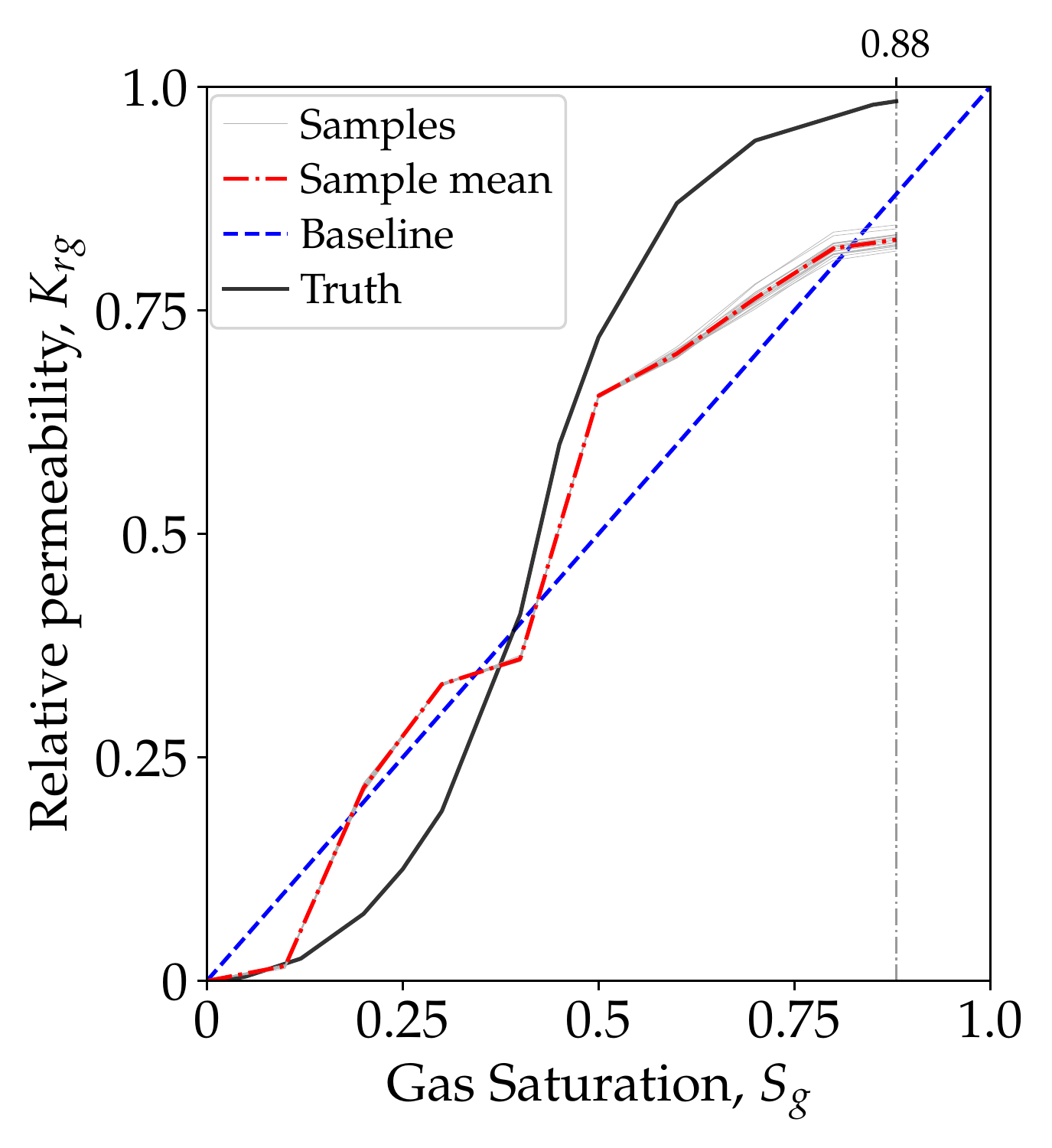}}
\hspace{0.5em}
\subfloat[Gas saturation field comparison]
{\includegraphics[height=0.36\textwidth]{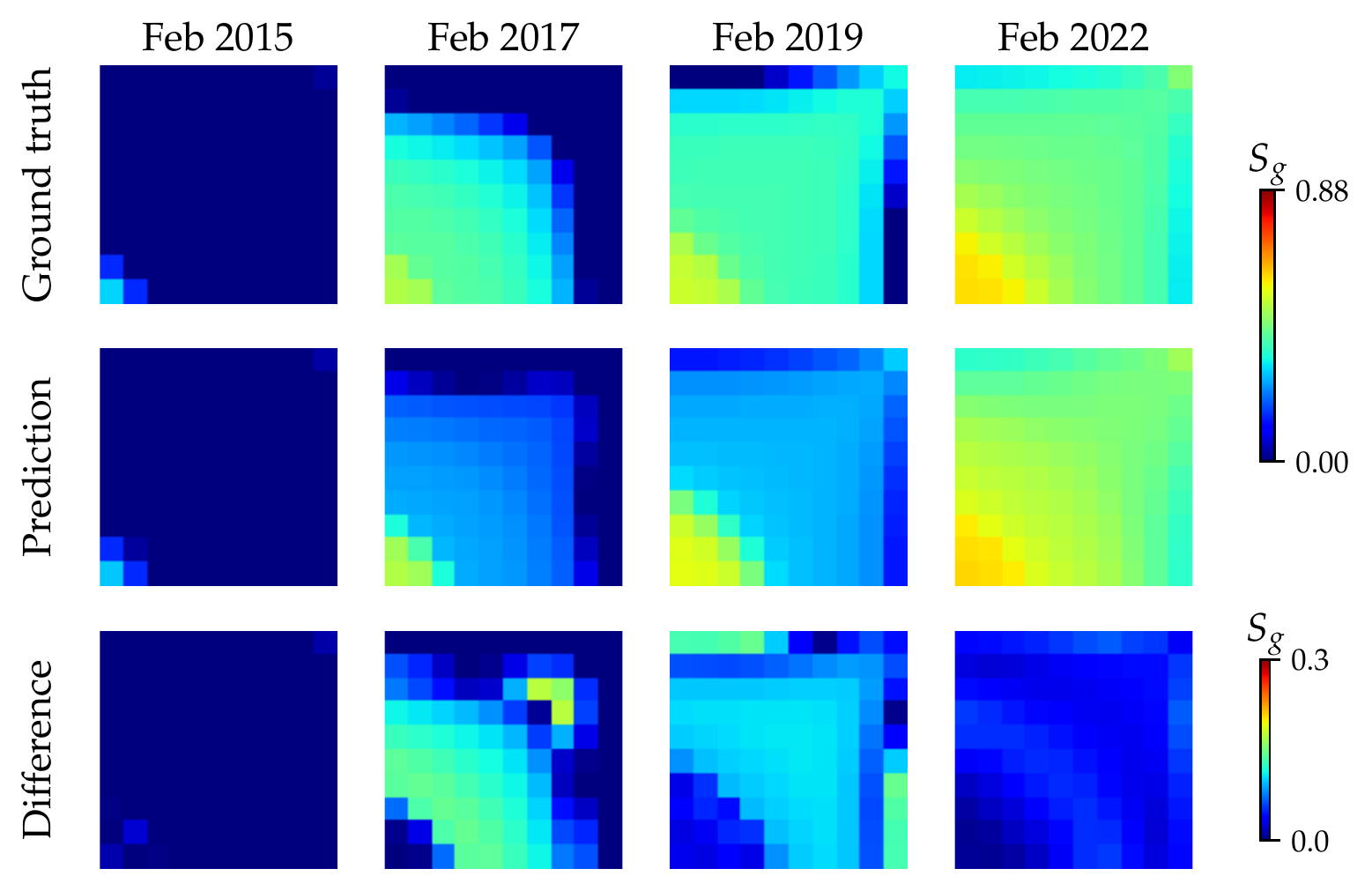}}\\
\subfloat[Well response comparison]
{\includegraphics[width=0.88\textwidth]{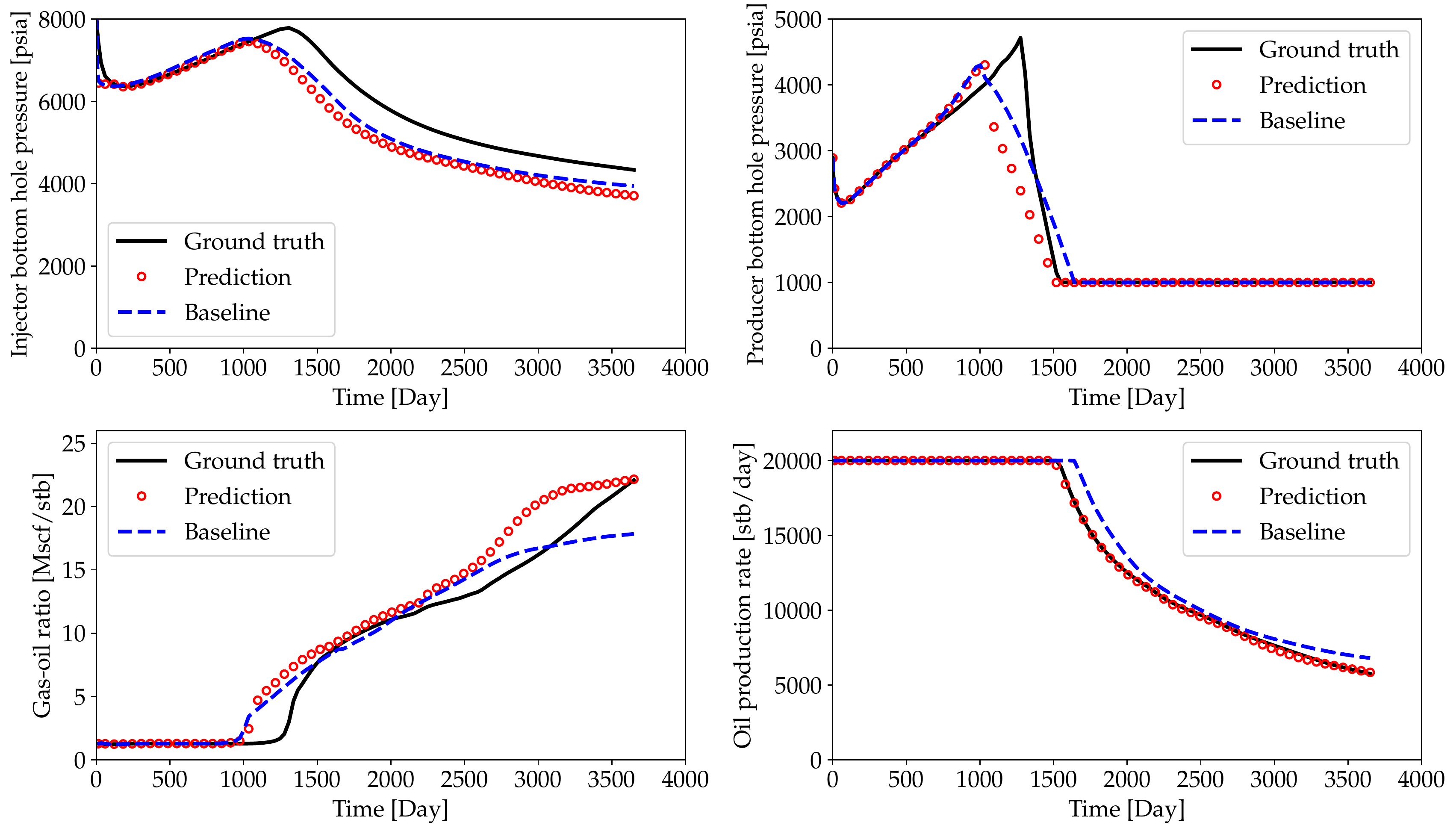}}
  \caption{Evaluation of the inferred relative permeability curve in SPE 1 benchmark: (a) comparison of the inferred curve with the ground truth, (b) comparison of the simulated gas saturation fields at four distinct years based on the inferred and true curves, and (c) comparison of the time series of well responses over 10 years based on the baseline, inferred, and true curves. Here, the oil production rate time series for all 10 years are used as the observation data.
  \label{fig:oildata-120}
  }
\end{figure*}

In contrast to the first experiment, the second experiment infers the relative permeability curve using a 10-year time series of oil production rate as the observation. The ill-posedness is reduced by using data in a longer time period. 
Nonetheless, the inferred relative permeability curve still differs from the true curve (Fig.~\ref{fig:oildata-120}a), although the predicted oil production rates are consistent with the ground truths (Fig.~\ref{fig:oildata-120}c). The predicted gas saturation fields and the other three well responses are significantly different from the corresponding ground truths, as illustrated in Figs.~\ref{fig:oildata-120}b and c, respectively.

From the above results, we can observe that the relative permeability curve may not be uniquely determined by the time series of production data alone. With sparse saturation data, the ill-posedness can be regularized and the estimation of relative permeability curve can be significantly improved.

\end{appendices}

\clearpage
\clearpage

\end{document}